\documentclass[]{aastex631}

	\usepackage{graphicx}	
	\usepackage{amsmath}
	\usepackage{amssymb}	
	\usepackage{longtable}
	\usepackage{threeparttable}
	\usepackage{multirow}
	\usepackage{booktabs}
	\usepackage{array}
    \usepackage[figuresright]{rotating}
    \usepackage[T1]{fontenc}   

     \received{January 26, 2025}
     \revised{February 11, 2025}
     \accepted{February 17, 2025}   

	\shorttitle{Superoutbursts and Positive Superhumps in AT Cnc}
	\shortauthors{Sun et al.}

	\graphicspath{{./}{figures/}}

\begin{document}
			\title{Superoutbursts and Positive Superhumps Occurred During the Standstill of a Z Cam-type Dwarf Nova}


\correspondingauthor{Sheng-Bang Qian}
\email{qiansb@ynu.edu.cn}	
\author{Qi-Bin Sun}
\affiliation{Department of Astronomy, School of Physics and Astronomy, Yunnan University, Kunming 650091, China}
\affiliation{Key Laboratory of Astroparticle Physics of Yunnan Province, Yunnan University, Kunming 650091, China}

\author{Sheng-Bang Qian}

\affiliation{Department of Astronomy, School of Physics and Astronomy, Yunnan University, Kunming 650091, China}
\affiliation{Key Laboratory of Astroparticle Physics of Yunnan Province, Yunnan University, Kunming 650091, China}

\author{Li-Ying Zhu}
\affiliation{Yunnan Observatories, Chinese Academy of Sciences, Kunming 650216, China}
\affiliation{University of Chinese Academy of Sciences, No.19(A) Yuquan Road, Shijingshan District, Beijing, China}

\author{Qin-Mei Li}
\affiliation{Department of Astronomy, School of Physics and Astronomy, Yunnan University, Kunming 650091, China}
\affiliation{Key Laboratory of Astroparticle Physics of Yunnan Province, Yunnan University, Kunming 650091, China}

\author{Fu-Xing Li}
\affiliation{Department of Astronomy, School of Physics and Astronomy, Yunnan University, Kunming 650091, China}
\affiliation{Yunnan Observatories, Chinese Academy of Sciences, Kunming 650216, China}

\author{Min-Yu Li}
\affiliation{Yunnan Observatories, Chinese Academy of Sciences, Kunming 650216, China}
\affiliation{University of Chinese Academy of Sciences, No.19(A) Yuquan Road, Shijingshan District, Beijing, China}	

\author{Ping Li}
\affiliation{Yunnan Observatories, Chinese Academy of Sciences, Kunming 650216, China}
\affiliation{University of Chinese Academy of Sciences, No.19(A) Yuquan Road, Shijingshan District, Beijing, China}	


			\begin{abstract}

Dwarf novae are semi-detached binaries, where a white dwarf accretes material from a cool main-sequence companion via an accretion disk, and are known for their intermittent outbursts, making them key systems for studying accretion physics. The accumulation of large survey datasets has challenged traditional models, which assumed that the disk remains hot and cannot produce superoutbursts during the standstill of Z Cam-type dwarf nova and that superoutbursts require a mass ratio of \( q = M_2/M_1 \leq 0.25 - 0.33 \). Here we report the detection of superoutbursts and positive superhumps (PSHs) during a standstill in the Z Cam-type star AT Cnc with a mass ratio larger than 0.33. Notably, the PSHs evolve gradually before the superoutburst begins, suggesting that an eccentric, precessing disk forms first, with the superoutburst occurring as the disk radius continues to expand. These findings provide the first detailed observational evidence of superoutbursts and PSHs occurring during standstill, offering important new insights into the classification of dwarf novae and the underlying mechanisms of outbursts.

\end{abstract}



\section{Introduction} \label{sec:intro}	
Dwarf novae (DNe) are a subclass of cataclysmic variable stars (CVs) characterized by intermittent outbursts \citep{warner1995cat}. 
The disk thermal-viscous instability model (DIM), widely used to explain these outbursts, suggests that viscous processes driven by magnetorotational instability (MRI) are key for angular momentum and mass transport in the accretion disk \citep{1974PASJOsaki, lasota2001disc}. 
During quiescence, the disk is in a low-temperature, low-viscosity state. As temperature rises to $\sim 8000$ K, partially ionizing hydrogen and triggering an instability. This leads to a transition to a fully ionized, high-temperature, high-viscosity state as the temperature continues to increase \citep{lasota2001disc, 2020AdSpR..66.1004H}. 
Z Camelopardalis (Z Cam) stars, a subclass of DNe, exhibit a unique ``standstill'' during outburst decline. This phenomenon is thought to result from the combined effects of the DIM and fluctuations in the mass transfer rate from the secondary star \citep{Meyer1983A&A, Smak1983ApJ}. The model suggests that in Z Cam, the mass transfer rate is typically near a critical value ($\dot{M}_{\text{crit}}$), and small deviations from this rate can trigger a standstill, corresponding to the hot branch in the DIM.
SU Ursae Majoris (SU UMa) is another subclass of DNe, are characterized by superoutbursts that are typically 0.5–1 mag brighter than normal outbursts and can last 5–10 times longer. These superoutbursts are accompanied by positive superhumps (PSHs), which are slightly longer than the orbital period. The thermal-tidal instability (TTI) \citep{osaki1985irradiation, whitehurst1991superhumps} extends the DIM framework to explain these phenomena. In the TTI model, mass transfer from the secondary star is assumed to be stable, and disk instabilities lead to the expansion of the disk beyond the 3:1 resonance radius, triggering a tidal instability that results in superoutbursts and PSHs. Tidal instability is expected in low-mass systems ($q = M_2 / M_1 \leq 0.25-0.33$), where the 3:1 resonance radius is smaller than the tidal truncation radius, limiting disk expansion \citep{Paczynski1977, 1989PASJ...41.1005O, 2021PASJ...73.1209W}.

AT Cancri (AT Cnc), a classic Z Cam-type DN, has been extensively studied, with multiple outbursts and standstills recorded \citep{Gotz1991, 1999PASJ...51..115N, 2001IBVS.5099....1K, 2004A&A...419.1035K}. The orbital period was initially determined to be 0.2011(6) days from spectroscopic observations \citep{1999PASJ...51..115N}. More recent photometric data have updated the orbital period to 0.201634(5) days \citep{2019NewA...67...22B}. Additionally, a classical nova shell surrounding the system was confirmed, likely resulting from a nova eruption $\sim$ 330 years ago \citep{2012ApJ...758..121S, 2017MNRAS.465..739S}.
This paper investigates the standstill, outbursts, and PSHs of AT Cnc using data from the Transiting Exoplanet Survey Satellite (TESS), All-Sky Automated Survey for Supernovae (ASAS-SN), Zwicky Transient Facility (ZTF), Wide Angle Search for Planets (SuperWASP), and the American Association of Variable Star Observers (AAVSO) (see Appendix and Fig. \ref{fig:alllight}). While recent studies have suggested the occurrence of superoutbursts in Z Cam systems \citep{2021PASJ...73.1280K,2019PASJ...71L...1K}, detailed variations were not resolved due to limitations in data resolution. For the first time, we present detailed evidence of superoutburst and PSH production during a standstill of AT Cnc, with PSHs occurring at least 22 days prior to the superoutburst. These results imply that the accretion disk of AT Cnc (\( q > 0.33 \)), exceeded both the tidal truncation radius and the 3:1 resonance radius during the standstill, continuing to expand even after surpassing the 3:1 resonance. Our analysis further shows that superoutbursts are triggered by oscillations, a characteristic feature of certain dwarf nova outbursts, and evolve from normal outbursts, suggesting that some parts of the disk may not reach a fully hot state during the standstills.
Furthermore, the PSH amplitude was found to be correlated with special outbursts, revealing that the localized thermal instability had a significant impact on the evolution of the eccentric disk. These findings suggest that superoutbursts are not exclusive to SU UMa-type DNe, challenge the conventional accretion disk model, and offer crucial observational evidence for investigating the origins of standstills, PSHs, and superoutbursts.

\section{Results}\label{sec:Results}

This study focuses on the TESS photometric data, supplemented by additional observations from other projects. 
TESS observed two short outbursts of AT Cnc (Sector 21: S21) and two standstills (S44-S47 and S71-S72; the top panel of Fig. \ref{fig:lc_FT}). Frequency analysis of the S21, S44-S47, and S71-S72 segments was conducted using Period04 software \citep{Period04}. In S21, there is a weak signal at twice the orbital frequency (the bottom panel of Fig. \ref{fig:lc_FT}(a)). Stable orbital signals are also present in the S44-S47 and S71-S72 segments, with an average orbital period of 0.201754(3) days (4.9565(8) d$^{-1}$). Quasi-periodic variations were observed during S44-S47 and S71, with mean periods of P$\rm _{S44-S47}$ = 5.577(4) days (Amplitude$\rm _{S44-S47}$ $\sim$ 0.03 mag) and P$\rm _{S71}$ = 5.056(3) days (Amplitude$\rm _{S71}$ $\sim$ 0.11 mag). These periods are shorter than the outburst cycle ($ \sim $ 9.53 days in V-band, see Appendix). Similar quasi-periodic behavior was observed during two other standstill intervals: JD2457742.3908 to JD2457792.4472 (Fig. \ref{fig:standstill}(b) and (e)), and JD2458152.8375 to JD2458213.9356 (Fig. \ref{fig:standstill}(c) and (f)), with mean periods of 6.43(5) days and 8.08(1) days, respectively. 

Two types of precessing disks are found in CVs: the eccentric prograde precessing disk and the tilted retrograde precessing disk. Tilted retrograde precessing disk induces a superorbital signal, while interactions between the mass steams from the secondary star and the tilted disk generate negative superhumps (NSHs), which are shorter than the orbital period \citep{Barrett1988MNRAS, wood2009sph}. Recent studies have provided evidence for the existence of tilted disks and NSHs \citep{Sun2023arXiv, Sun2024ApJ, 2024arXiv240704913S}.
No NSHs were detected in the S44-S47 and S71 (Fig. \ref{fig:lc_FT}), excluding the possibility of tilted disk precession. The predicted precession period of an eccentric disk is $\sim$1.82days($\rm 1/P_{prec} = 1/P_{orb} - 1/P_{psh}$), ruling out luminosity fluctuations from prograde precession of an eccentric disk. Between JD2457425.5479 and JD2457536.8826, AT Cnc exhibited an interesting behavior: following a brightness drop of about 0.7 mag after an outburst, a standstill lasting approximately 15 days occurred. At the end of this standstill, AT Cnc failed to fully quiescence and instead entered a new outburst cycle (Fig. \ref{fig:standstill}(g)). This new outburst lasted about 8 days before transitioning into a phase of repetitive oscillations, with amplitude and period gradually decreasing before reaching a minimum, after which they increased again, eventually evolving into a typical outburst pattern. The shortest oscillation interval was about 5 days, similar to the oscillations observed during S44-S47 and S71. We conclude that the oscillations observed during S44-S47 and S71 likely represent special DN outbursts, which evolve from normal outbursts, decreasing in period and amplitude until reaching standstill, and then resume as normal outbursts following the standstill. The oscillations in S71 show a general increasing trend, which eventually triggers the superoutburst, further confirming that these oscillations are a characteristic feature of special DN outbursts. The superoutburst is approximately 0.7 mag brighter than the special outburst, lasts at least 30 days, and is accompanied by significant PSHs (Fig. \ref{fig:psh1}), similar to SU UMa-type DNe.

PSHs were detected in both S71 and S72, and detailed analysis of their evolution was performed using locally-weighted polynomial regression (LOWESS), segmented Fourier analysis, Continuous Wavelet Transform (CWT), and Gaussian fitting (see Appendix), obtaining the following results:

\begin{enumerate}
	\item[(a)] PSHs consistently emerge regardless of whether the data is divided into two-day or three-day segments for segmented frequency analysis (Fig. \ref{fig:2-3_FT}). Averaging the results from the three-day segments gives an average PSH period of 0.22689(59) days and amplitude of 0.048(1) mag. The value of $\epsilon$ is determined to be 0.1246(31) ($\epsilon = (P_{\text{psh}} - P_{\text{orb}}) / P_{\text{orb}}$). Analysis of the frequency spectra, CWT, and amplitude variations between maxima and minima shows that the PSH amplitude increases during S71, peaks at the superoutburst during S72 (Fig. \ref{fig:psh1}).
	
	\item[(b)] A comparative analysis of the residuals (residual$_{2}$) from the light curve during S71, segmental Fourier transform results, and CWT shows that PSHs do not evolve linearly. As the overall brightness increases, the PSH amplitude exhibits oscillatory growth, with amplitude variations corresponding to the special DN outbursts. Specifically, the PSH amplitude increases as the special DN outbursts weaken and decreases when the outbursts intensify (Figs. \ref{fig:psh1} and \ref{fig:S71_O-C}).
	
	\item[(c)] The segmented frequency analysis reveals that the PSH period gradually increases during S71 and decreases after the onset of the superoutburst. This trend is confirmed by the O-C analysis. A parabolic fit to the entire O-C dataset gives a period change rate of $ 		
	\dot{P} = -1.74 (\pm 0.07) \times 10^{-4} \, \text{days/day} $
	The evolution of the PSH period in SU UMa-type DNe is divided into three stages \citep{2009PASJ...61S.395K,Kato2015PASJ}: Stage A (early evolution, long period), Stage B (positive period derivative), and Stage C (shorter, stable period). Stage B is found to be in  AT Cnc, with the PSH period consistently decreasing during Stage A and C (Figs. \ref{fig:psh1}(g) and \ref{fig:S71_O-C}). The parabolic fit for Stage B yields a period change rate of $ \dot{P} = +1.77 (\pm 0.84) \times 10^{-4} \, \text{days/day} $.
	
\end{enumerate}

\section{DISCUSSION} \label{sec:DISCUSSION}
\begin{figure}
	\centering
	\includegraphics[width=\columnwidth]{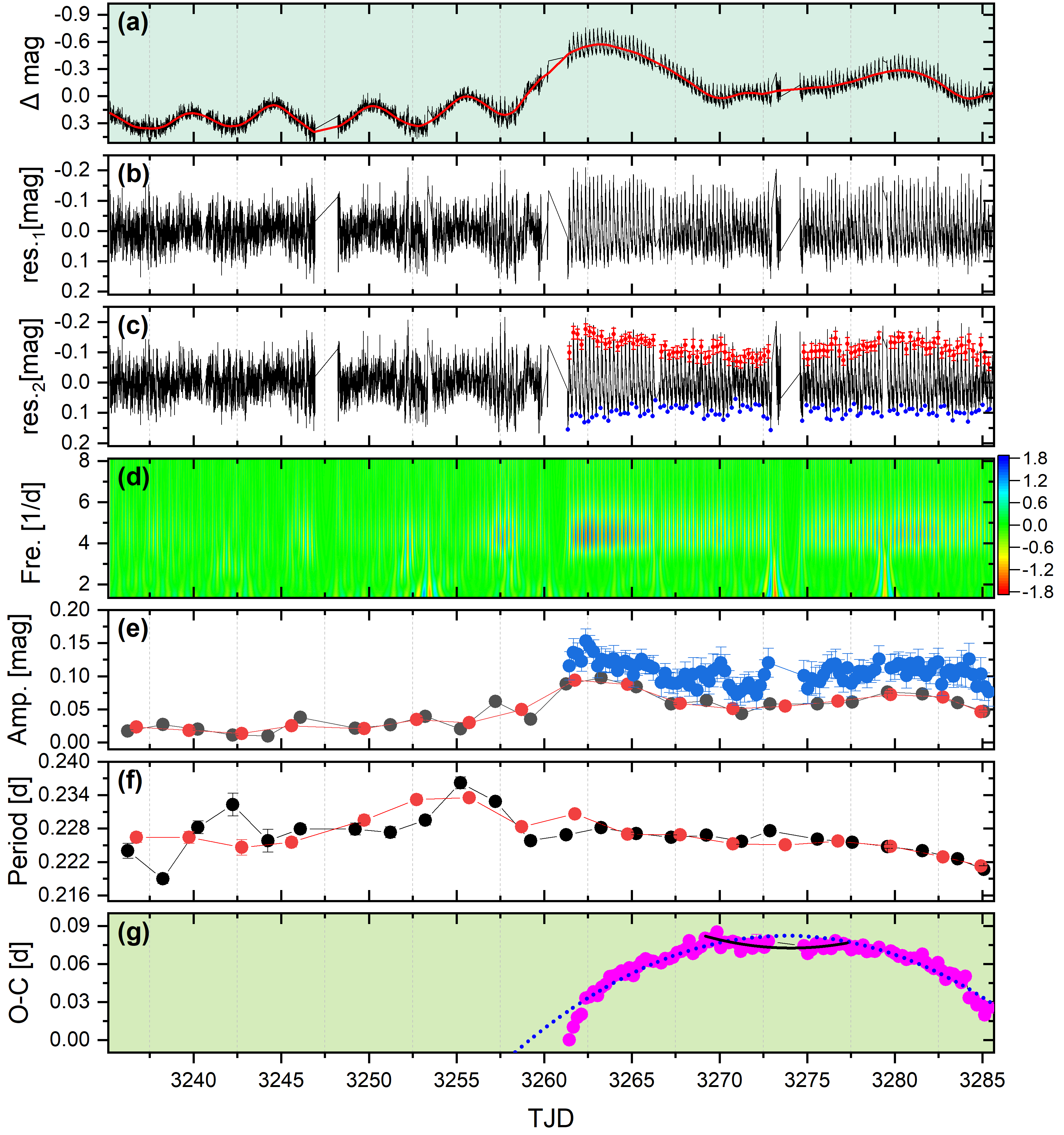}
	\caption{Results of PSH analysis in AT Cnc. (a) Light curves of S71-S72 from TESS with the LOWESS fit (red solid line); 
		(b) Residuals from the LOWESS fit (residual$_{1}$); 
		(c) Light curve after removing the orbital signal (residual$_{2}$), with red and blue dots representing PSH maxima and minima, respectively; 
		(d) 2D power spectra of the CWT for residual$_{2}$, with colors indicating oscillation strength; 
		(e) Amplitudes from Fourier analysis of 2-day (black), 3-day (red) segments, and the difference between maxima and minima (blue); 
		(f) PSH periods derived from Fourier analysis of 2-day (black) and 3-day (red) segments; 
		(g) O-C analysis of PSH maxima, with blue dotted and black solid lines representing parabolic fits to all O-C values and Stage B, respectively.}
	\label{fig:psh1}
\end{figure}

The standstill phenomenon in Z Cam systems is generally attributed to the combination of the DIM and fluctuations in mass transfer. Statistical results and simulations support this explanation \citep{Meyer1983A&A, dubus2018testing, 2020AdSpR..66.1004H}. However, the recent discovery of a unique Z Cam-type DN, specifically an IW And-type object \citep{Simonsen2011AAS, Kato2019PASJ}, has posed significant challenges to this established model, as it demonstrates that the standstill can unexpectedly terminate in an outburst.
 According to DIM, the accretion disk during the standstill phase is expected to be in a hot state, transitioning into quiescence. In contrast, IW And systems often exhibit an outburst after standstill, deviating from this behavior. One hypothesis suggests that secondary mass-transfer outbursts may drive accretion disk instability, though the origin of these outbursts remains unclear \citep{2014mass-transferoutbursts}. Alternatively, a tilted-thermal instability model (tilted-DIM) proposes that the inner tilted disk maintains its hot state by allowing mass to flow inward, while the outer disk remains thermally unstable \citep{kimura2020thermal}. However, simulations based on the tilted-DIM struggle to fully reproduce the details IW And phenomenon \citep{kimura2020thermal, 2024arXiv240903011S}.
In this study, we analyze TESS data and identify distinct long-period oscillations during the standstill phase of AT Cnc. These oscillations evolve from normal outbursts and transition back to outbursts as the standstill ends (Fig. \ref{fig:standstill}(g)). The amplitude of the oscillations in S71 increases gradually, ultimately triggering a superoutburst (Fig. \ref{fig:psh1}), resembling those observed in SU UMa systems. This behavior suggests that the oscillations represent a special type of DN outburst. Our findings imply that both superoutbursts and normal DN outbursts can occur during the standstill, challenging the traditional DIM framework. Specifically, during some Z Cam standstill phases, as proposed by the tilted-DIM, the disk may not be in a fully hot state but may instead exhibit a localized high-temperature zone. The absence of a tilted disk in AT Cnc suggests that this incomplete hot state may not directly correspond to the tilted disk.

\begin{figure}
	\centering
	\includegraphics[width=0.45\columnwidth]{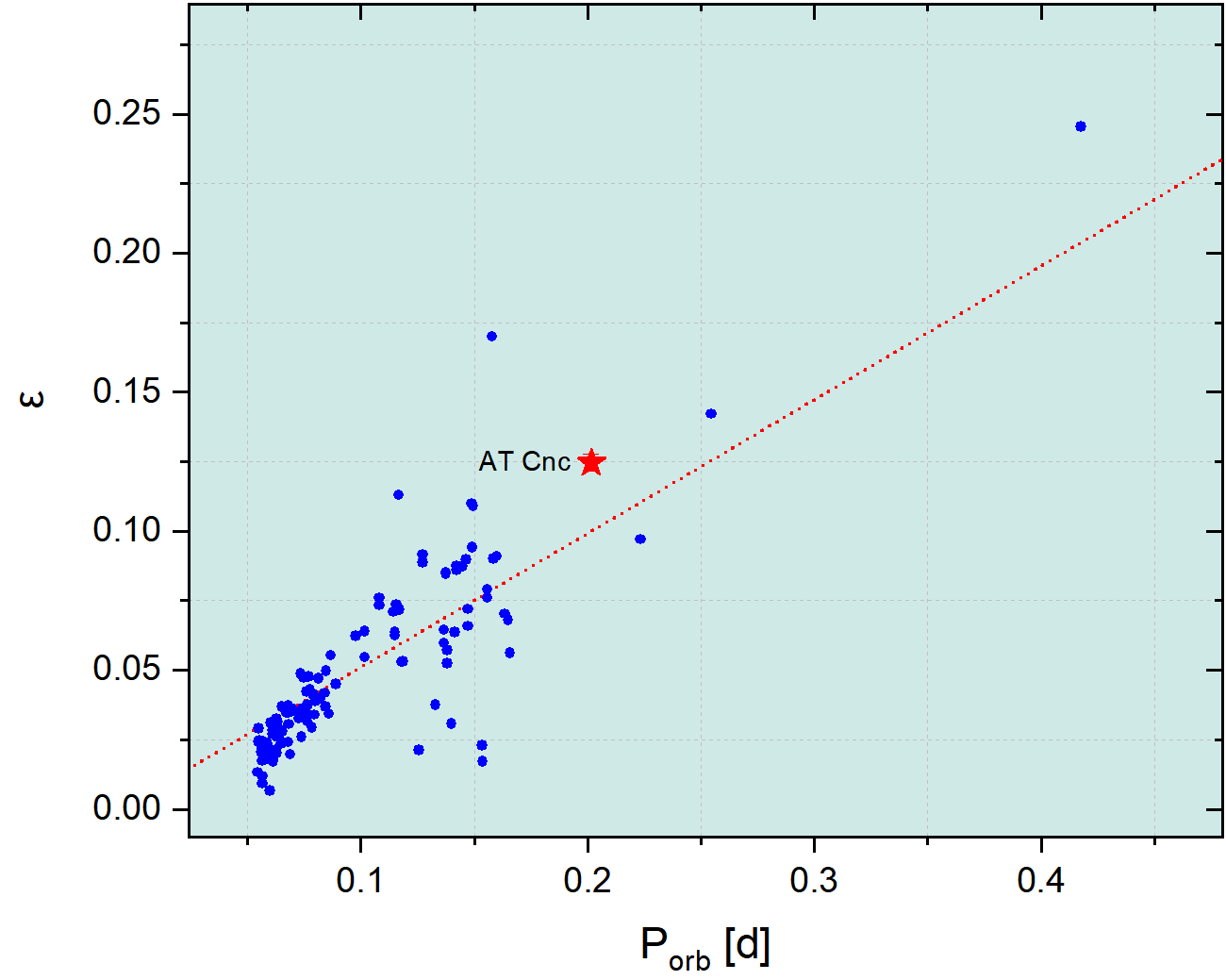}
	\includegraphics[width=0.45\columnwidth]{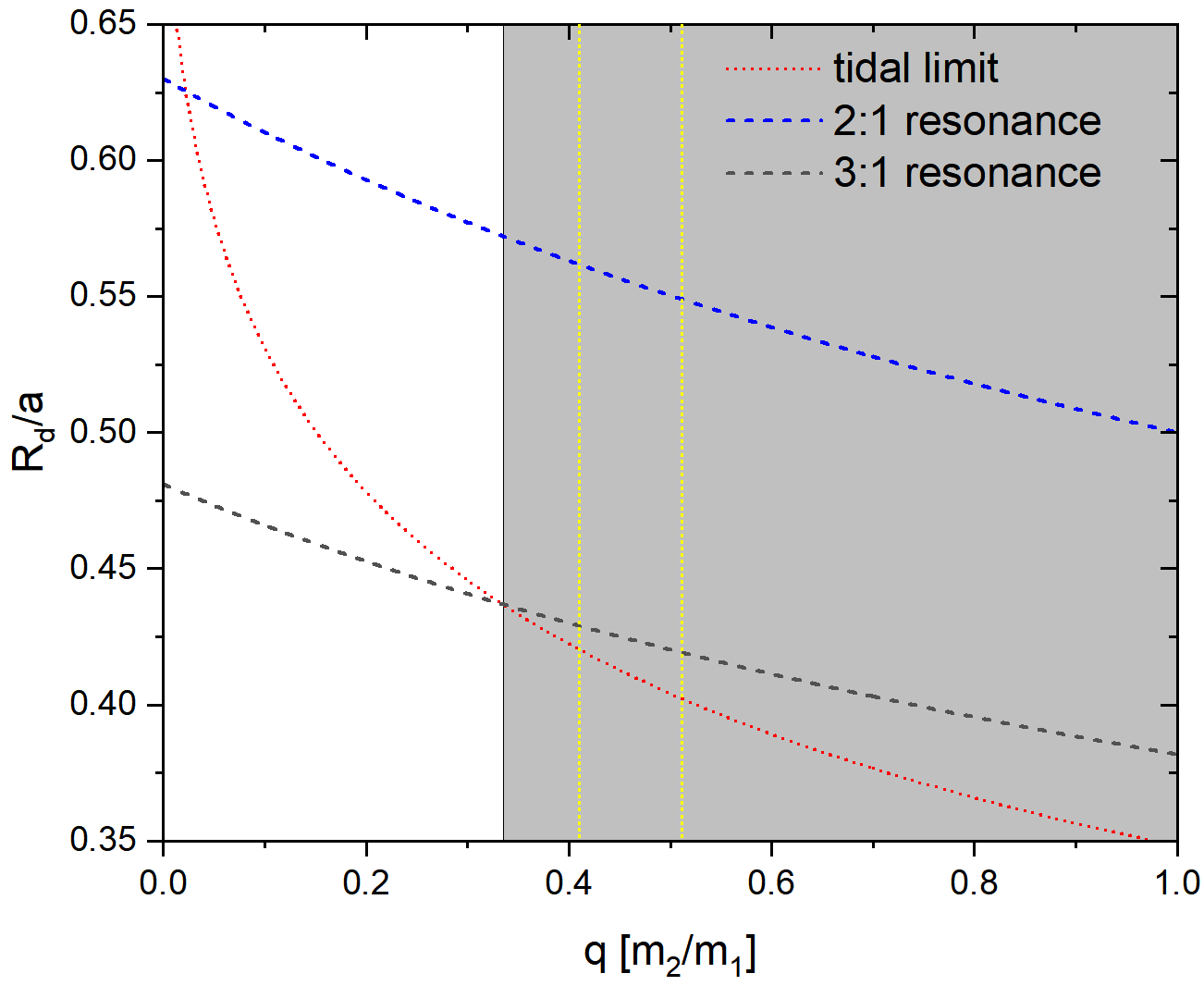}
	\caption{Left panel: blue dots from the literature \citep{BruchIII,2005PASP..117.1204P}, red dashed line is P$\rm _{orb} $- $\epsilon$ relation \citep{BruchIII}, red pentagram corresponds to the position of AT Cnc. Right panel: black, blue and red dotted lines correspond to 3:1 resonance radius, 2:1 resonance radius and tidal truncation radius, vertical yellow dashed lines correspond to $ q=0.41 $ and $ q=0.511 $. The gray area corresponds to mass ratios greater than 0.33 and the 3:1 resonance radius is larger than the tidal truncation radius. \label{fig:q} }

\end{figure}

The TTI model suggests that when the disk expands beyond the 3:1 resonance radius, tidal instability induces the formation of an eccentric disk, leading to superoutbursts and the development of PSHs \citep{1989PASJ...41.1005O}. Upon reaching the tidal truncation radius, tidal torque extracts angular momentum, halting further disk expansion \citep{Paczynski1977,2021PASJ...73.1209W}. The model predicts that systems exhibiting this behavior typically have mass ratios below 0.25 \citep{1988MNRAS.232...35W,2005PJABOsaki}, with extreme cases below 0.33 \citep{2000NewAR..44...51M}. Recent studies using TESS data derived an empirical relationship between the orbital period and $\epsilon$ (P$_{\rm orb}$ - $\epsilon$) \citep{BruchIII}. The $\epsilon$ value for AT Cnc aligns with this correlation (Fig. \ref{fig:q}, left panel). While the mass ratio of AT Cnc was initially estimated as 0.511 \citep{1999PASJ...51..115N}, this was re-estimated to 0.41 based on the empirical $\epsilon(q)$ relation \citep{2005PASP..117.1204P}. Using methods similar to previous studies \citep{2003A&A...401..325O,2008PASJ...60L..23K}, we plotted the relationship between the 3:1 and 2:1 resonance radii and the tidal truncation radius (Fig. \ref{fig:q}, right panel). For both mass ratios (0.511 and 0.41), the 3:1 resonance radius exceeds the truncation radius. It is important to acknowledge that the mass ratio estimates based on PSHs are rough and more accurate estimates of the mass ratio of AT Cnc are needed. We suggest that radial velocity measurements are necessary in future work. SU UMa-type superoutbursts, all predicted to generate PSHs after the final normal outburst, have been classified into various types \citep{2003A&A...401..325O,2005PJABOsaki}. Observations confirm PSH formation after a superoutburst reaches a certain stage \citep{2019MNRAS.488.4149C,2023ApJ...954..135W}. PSHs were detected in AT Cnc from the start of S71, with a superoutburst triggered at least $ \sim $22 days after the the formation of an eccentric disk. This suggests the formation of an eccentric, prograde-precessing disk before significant tidal dissipation, with the accretion disk continuing to expand past both the truncation radius and the 3:1 resonance radius.

\begin{figure}
	\centering
	\includegraphics[width=0.45\columnwidth]{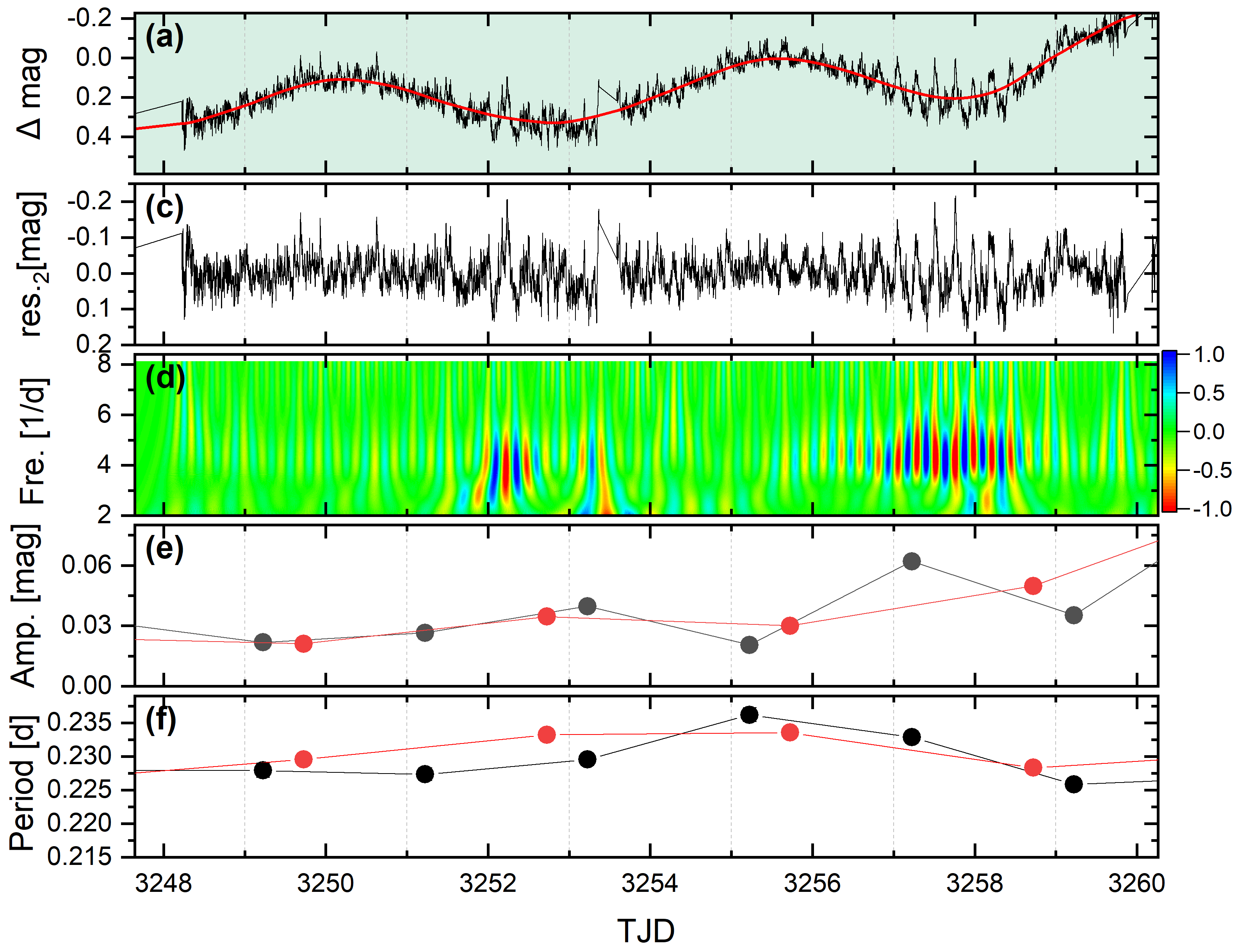}
	\includegraphics[width=0.45\columnwidth]{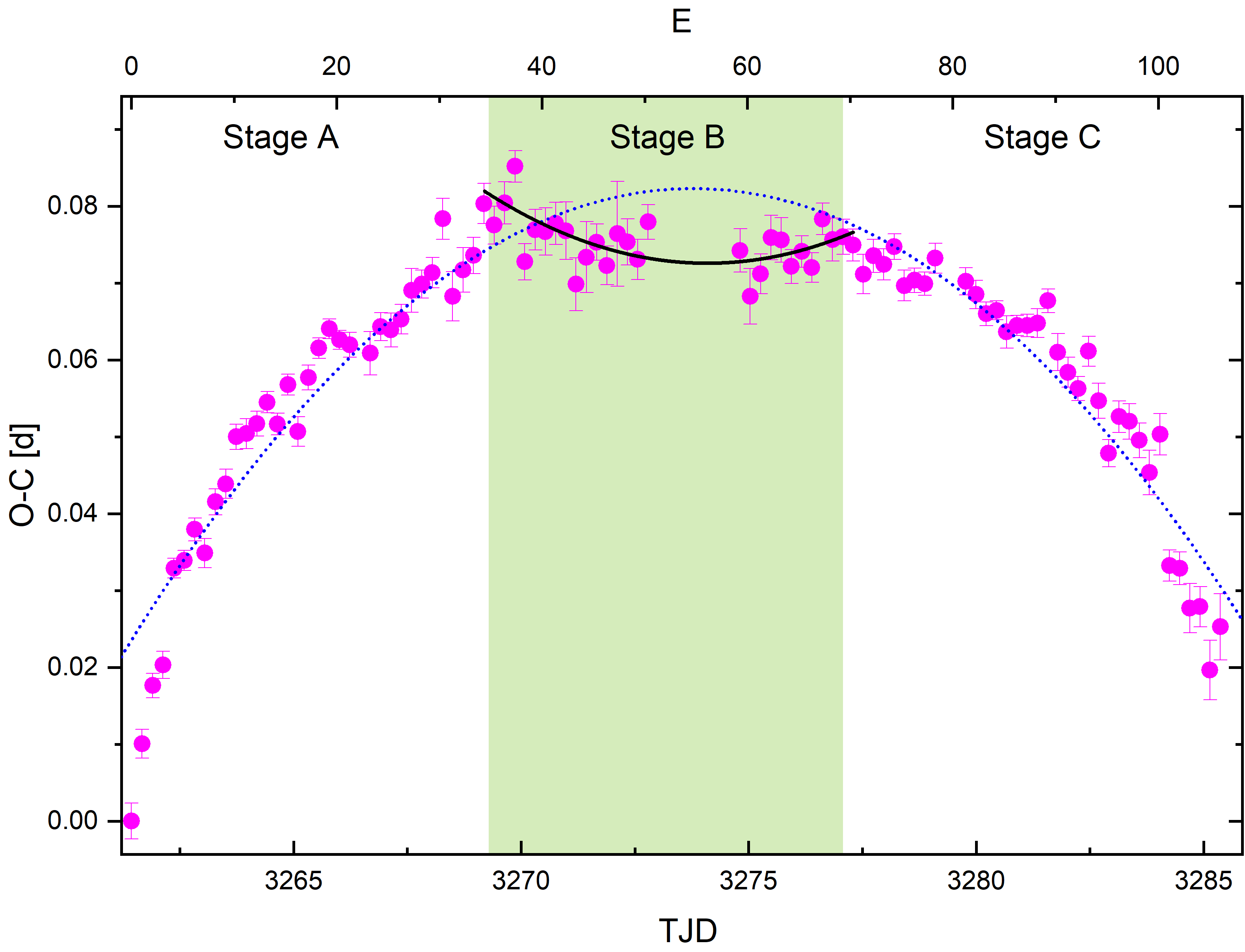}
	\caption{Left Panel: A magnified view of the changes in PSHs within S71 data from TESS, with individual panels mirroring the presentation style of Fig. \ref{fig:psh1}.
		Right Panel: The O-C plot of the maxima of PSHs in S72, which is consistent with the data representation of Fig. \ref{fig:psh1}(g).\label{fig:S71_O-C}}
\end{figure}

We observe that the PSH amplitude oscillates, increasing before the superoutburst. The amplitude rises as the special outbursts wane, reaching a peak at the bottom (Figs. \ref{fig:psh1} and \ref{fig:S71_O-C}). This is key evidence that special outbursts have had a significant impact on the evolution of eccentric disks. PSHs are attributed to periodic tidal forces exerted by the secondary star on the eccentric disk, particularly on its protruding portion. In the TTI framework, the accretion disk expands during normal outbursts and contracts during their decline. The larger disk radius at the peak of the special outbursts corresponds to greater tidal pressures, yet this appears contrary to the observed PSH amplitude changes in AT Cnc. Irradiation-induced mass transfer variations, where the secondary star is irradiated by the accretion disk during outbursts, which increases the mass transfer, may limit disk expansion \citep{1995AcA....45..355S,2000A&A...364L..75K,2020AdSpR..66.1004H}. In the low state, however, the disk continues to expand, which correlates with the observed PSH amplitude variations in AT Cnc. While the role of irradiation-induced mass transfer remains debated, the observed PSH amplitude changes may offer insights into the underlying mechanism. Furthermore, the rate of eccentricity growth should be considered \citep{1991ApJ...381..259L,2000MNRAS.314L...1M,2005PJABOsaki}. If eccentricity growth exceeds the rate of disk contraction during the low state, leading to an extended eccentric disk, this could explain the observed PSH amplitude variations in AT Cnc. However, further simulations and calculations are required to fully understand the specific physical processes involved.

\begin{figure*}
	\centering
	\includegraphics[width=0.75\columnwidth]{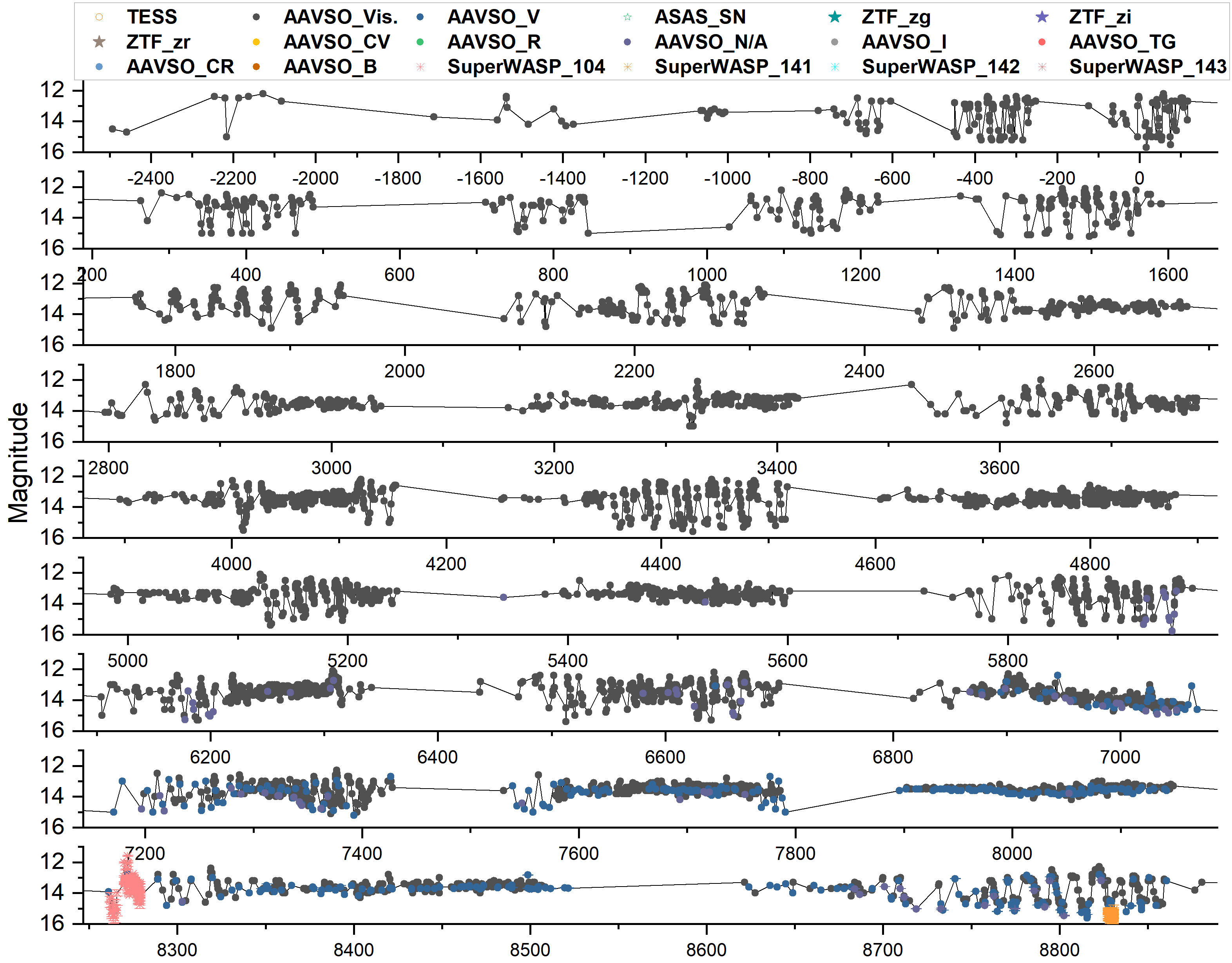}\\
	\includegraphics[width=0.75\columnwidth]{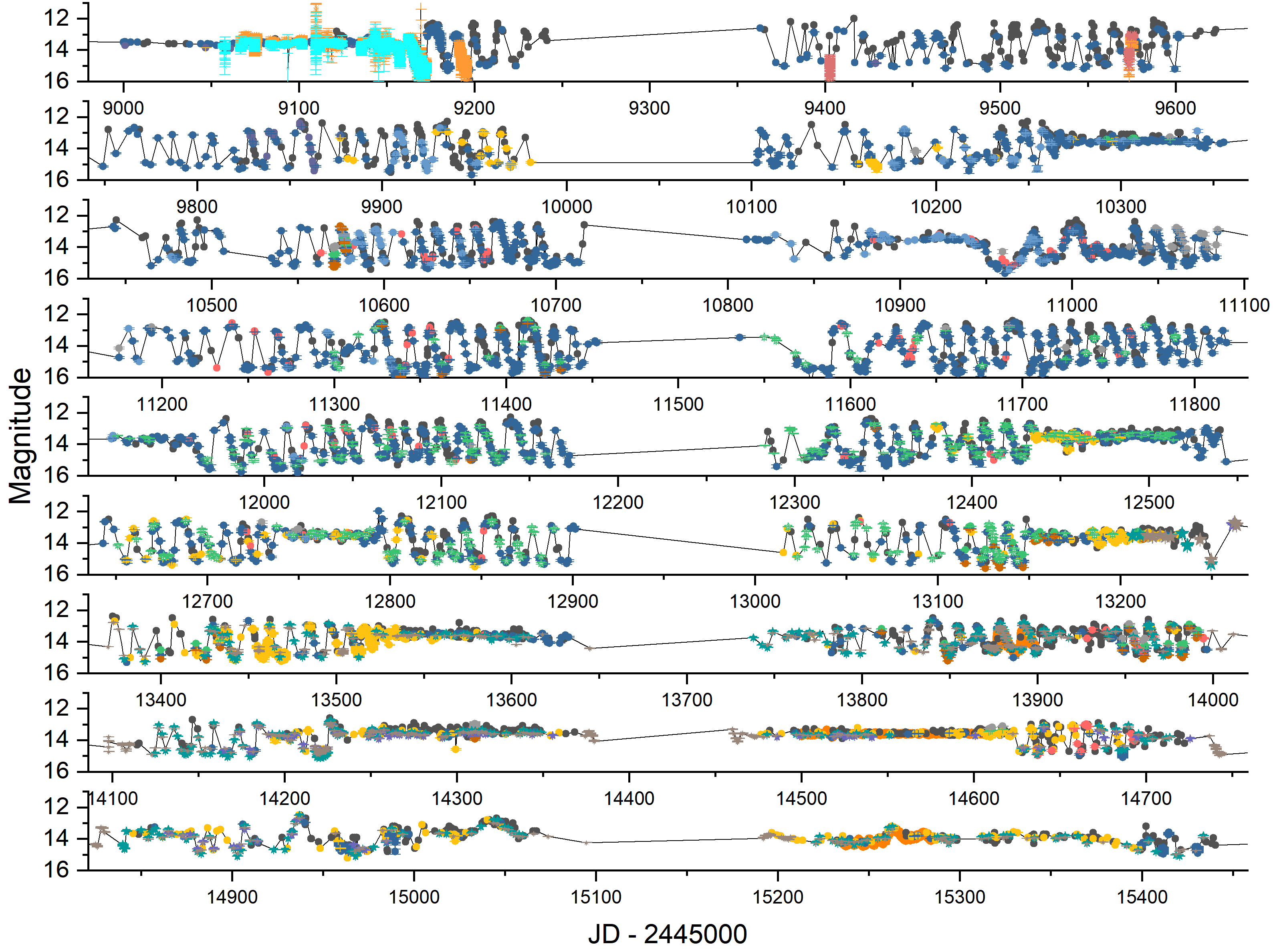}
	\caption{AT Cnc light curves for different data sources (see Section \ref{sec:Data}), refer to the figure legend for the meaning of the different symbol codes.\label{fig:alllight}}
\end{figure*}

\begin{figure*}
	\centering
	\includegraphics[width=0.7\columnwidth]{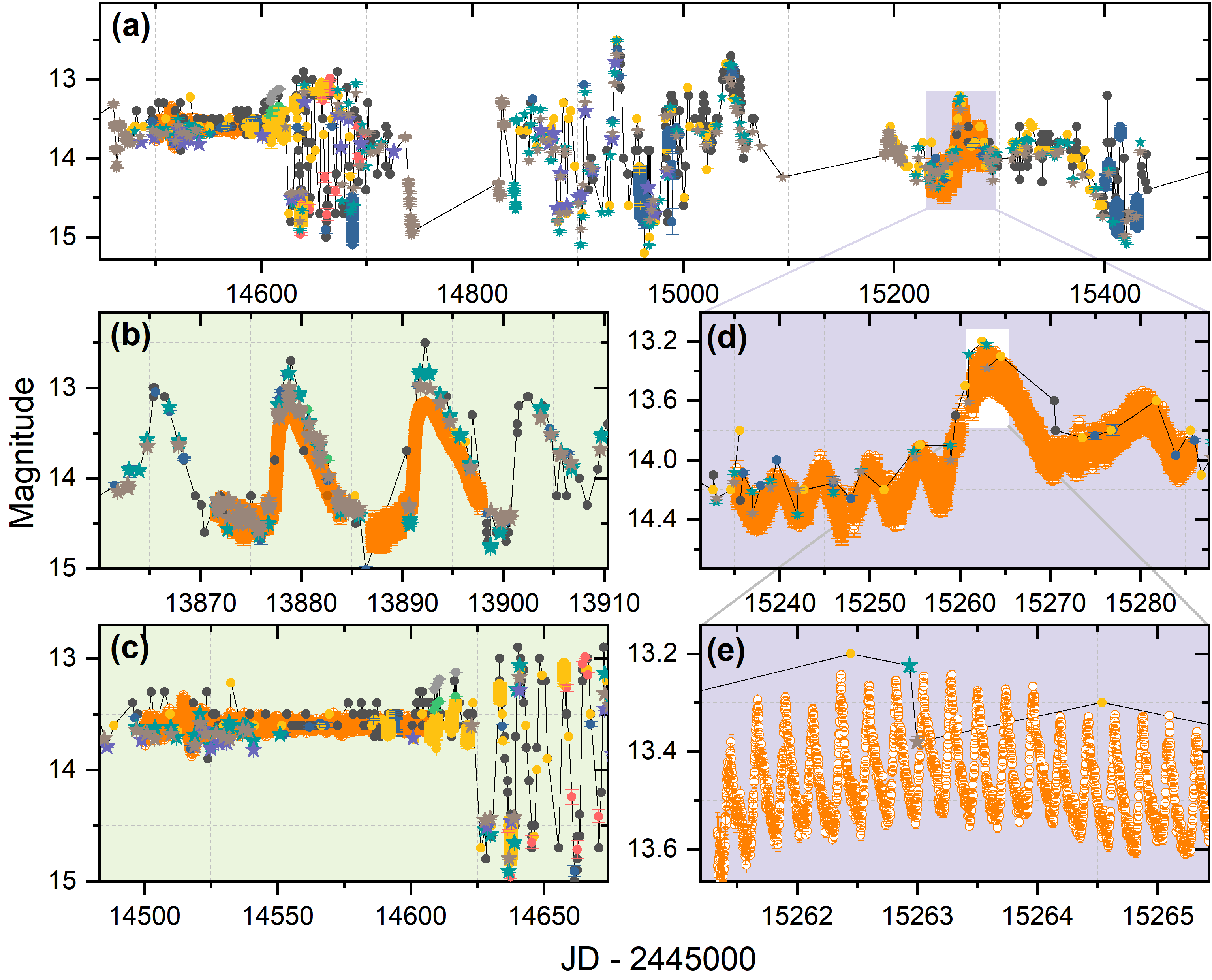}\\
	\includegraphics[width=0.7\columnwidth]{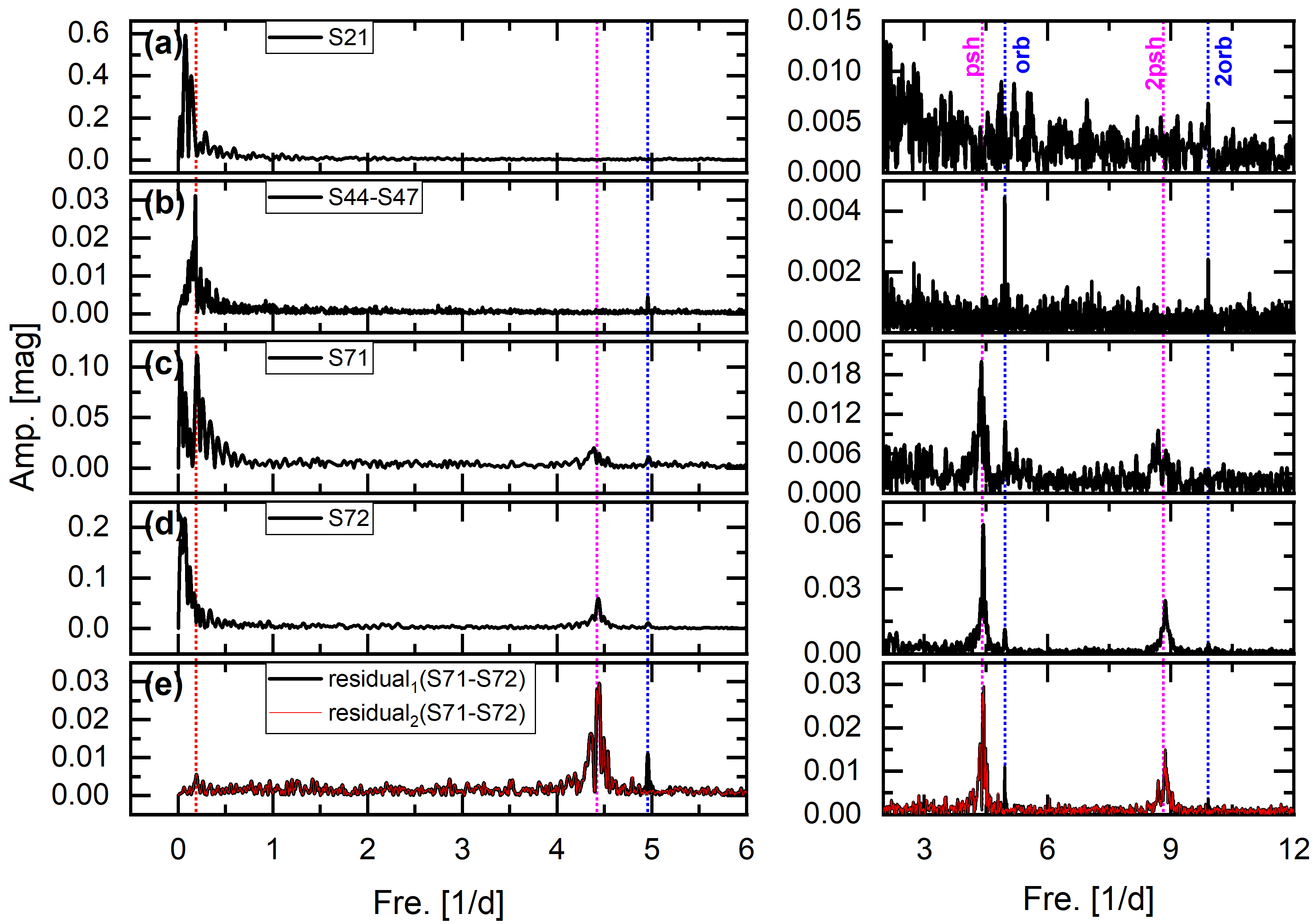}
	\caption{Top panels: (a) Light curve segment of AT Cnc from the localized magnification in Fig. \ref{fig:alllight}, the orange circle data comes from TESS; 
		(b) Two short outbursts observed during S21; 
		(c) Standstill during S44-S47; 
		(d) Oscillations and superoutbursts during S71-S72; 
		(e) Magnified view of the peak in panel (d).
		Bottom panels: Frequency spectra corresponding to S21, S44-S47, S71, S72, and S71-S72, corresponding to panels (a), (b), (c), (d), and (e), respectively. 
		The black and red curves in panel (e) correspond to panels (b) and (c) in Fig. \ref{fig:psh1}. The right section shows an enlarged view of the left side, with vertical dashed lines marking the oscillatory frequency, PSH frequency, orbital frequency, twice the PSH harmonics, and twice the orbital harmonics.}
	\label{fig:lc_FT}
\end{figure*}

\begin{figure}
	\centering
	\includegraphics[width=0.7\columnwidth]{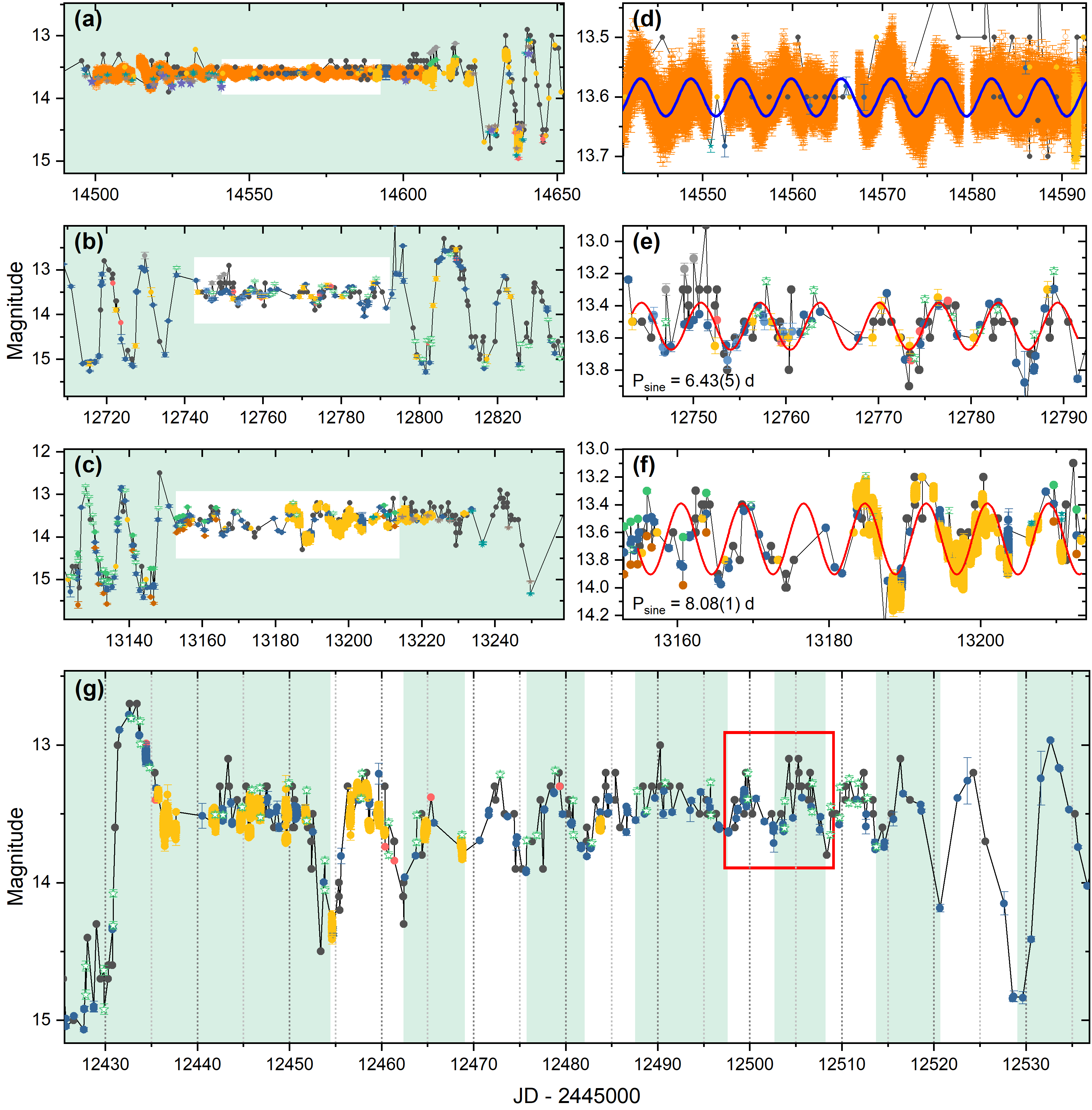}
	\caption{Oscillations during the brightness standstill of AT Cnc.
		Panels (a), (b), and (c) display three light curves showing brightness standstill with oscillations. Panels (d), (e), and (f) provide magnified views of the white rectangular regions from panels (a), (b), and (c), respectively, with sine fits overlaid on the curves.
		Panel (g) illustrates the oscillation generation process, with the red rectangular box highlighting the two shortest oscillation intervals, approximately 5 days. The data symbols in the figure are consistent with Figure \ref{fig:alllight}.\label{fig:standstill}}
\end{figure}

\section{Acknowledgements}

This work was supported by National Key R\&D Program of China (grant No. 2022YFE0116800), the National Natural Science Foundation of China (Nos. 11933008). We would like to thank the teams and facilities behind the following surveys for providing the data used in this study: the All-Sky Automated Survey for Supernovae for their V-band data on AT Cnc, the Zwicky Transient Facility for their long-term monitoring of AT Cnc in multiple filters, the SuperWASP survey for their observations of AT Cnc over several years, the American Association of Variable Star Observers for their extensive collection of variable star data, and the Transiting Exoplanet Survey Satellite for providing high-quality space-based photometric data on AT Cnc. Their contributions have been invaluable to this research. All of the TESS data used in this paper can be found through the MAST: \dataset[doi:10.17909/kydh-vx82]{https://doi.org/10.17909/kydh-vx82}

\section{Appendix}\label{sec:Appendix}
A description of the data used in this paper and an estimate of the outburst periods, as well as some figures and tables, are included in the Appendix.

\textbf{Figure \ref{fig:2-3_FT}}: Frequency spectra of segmented frequency analysis.
\textbf{Figure \ref{fig:outburst-FT}}: Analysis of the AT Cnc outburst periods.
\textbf{Table \ref{tessLOG}}: TESS photometric log for AT Cnc.
\textbf{Table \ref{tab:tab_FT}}: Frequency analysis results for different sectors and segments.

\subsection{Data Sources}\label{sec:Data}

The All-Sky Automated Survey for Supernovae (ASAS-SN) \citep{2019MNRAS.486.1907J} is an automated ground-based survey comprising 24 small telescopes, designed for rapid supernova detection across the entire sky, with a limiting magnitude of V $\sim$ 17. AT Cnc (ASAS-SN ID: ASASSN-V J082836.99+252002.8) was monitored by ASAS-SN over five years in the V band, with an average magnitude of 13.92. Data were retrieved from the ASAS-SN Variable Stars Database \footnote{https://asas-sn.osu.edu/variables}.

The Zwicky Transient Facility (ZTF) \citep{2019ZTF} is a time-domain optical survey using the Palomar 48-inch Schmidt Telescope, scanning the northern visible sky at 3,760 square degrees per hour, with median depths of $g \sim 20.8$ and $r \sim 20.6$. AT Cnc (ZTF ID: ZTF17aaagvsz) was observed by ZTF for approximately six years (MJD 58206.2148 to MJD 60433.1983), using ZTF filters ZTF\_zg (mean magnitude 13.86), ZTF\_zr (mean magnitude 13.90), and ZTF\_zg (mean magnitude 13.86). Data were retrieved from the Lasair database \footnote{https://lasair.roe.ac.uk/}.

The SuperWASP survey, comprising two robotic observatories—SuperWASP-North (La Palma) and SuperWASP-South (South Africa)—operates in the 4000–7000 \AA\ range with photometric accuracy better than 1\% for stars between magnitudes 8 and 11.5 \citep{2010A&A...520L..10B}. AT Cnc (SuperWASP ID: 1SWASP J082836.92+252003.2) was observed by four cameras (104, 141, 142, and 143) over 3.6 years, with an average magnitude of 13.84, as part of the WASP DR1 dataset \footnote{https://wasp.cerit-sc.cz/}.

The American Association of Variable Star Observers (AAVSO) maintains a comprehensive record of AT Cnc observations, collected by a global network of astronomers across nine photometric bands, including Visual (Vis.), Johnson V (V), Johnson B (B), and Cousins R (R). For further details on these bands, see the AAVSO documentation\footnote{https://dev-mintaka.aavso.org/bands-optical-data-aavso-international-database}. The AAVSO database contains 74,874 data points over 50 years, providing valuable insights for the study of AT Cnc. Data were retrieved from the AAVSO public database \footnote{https://www.aavso.org/LCGv2/}.

The Transiting Exoplanet Survey Satellite (TESS) \citep{Ricker2015journal} is a NASA mission launched in 2018, primarily focused on exoplanet detection. TESS orbits Earth in a 13.7-day elliptical orbit and observes the sky in 27-day sectors using four wide-field cameras. Although primarily designed for exoplanet studies, TESS has also contributed to the study of cataclysmic variables (CVs). We have conducted long-term CV studies using TESS data \citep{Sun2023MNRAS,2023arXiv230905891S,2024arXiv240704913S}. AT Cnc (TESS ID: TIC 3645653) was observed in sectors S21, S44–S47, and S71–S72 (Table \ref{tessLOG}). TESS data are managed by the Mikulski Archive for Space Telescopes (MAST), providing two types of light curves: Simple Aperture Photometry (SAP) and Pre-search Data Conditioning Simple Aperture Photometry (PDCSAP). PDCSAP corrects for sector discontinuities and removes systematic errors \citep{2016SPIE.9913E..3EJ}, though it can occasionally distort DN outburst light curves \citep{2023arXiv230911033S,2024arXiv240903011S}. Therefore, we used PDCSAP data for all sectors except S71–S72, where SAP data were used. TESS fluxes were converted to relative magnitudes using the equation $\Delta \text{mag} = -2.5 \log_{10} (\text{flux})$ \citep{Osaki2013b}. For comparison with other photometric data, 14 mag was added to the S21 and S71-S72 segments, while 13.6 mag was added to the S44-S47 segment. For subsequent detailed analyses, we have used the relative magnitudes. TESS data were retrieved from MAST \footnote{https://mast.stsci.edu/}.

We identify two short outbursts of AT Cnc during S21, with an interval of $\sim$ 13.6 days (top panel of Fig. \ref{fig:lc_FT}(b)). The outburst peaks brighter than 13 mag, while the quiescent state was $\sim$ 14.5 mag. Between S44 and S47, a standstill at $\sim$ 13.6 mag is observed. A fluctuating standstill around 14 mag is noted between S71 and S72, during which superoutbursts and PSHs were detected. Two outbursts occurred between S44-S47 and S71-S72, both exceeding 13 mag (top panel of Fig. \ref{fig:lc_FT}(a)). However, due to missing data, pre- and post-outburst variations cannot be estimated, preventing confirmation of whether these were superoutbursts. Due to space constraints, detailed discussion of each outburst and standstill is omitted. Frequency analysis of the V-band data reveals an outburst period of $\sim$ 9.53 days 
(Fig. \ref{fig:outburst-FT}). Sinusoidal fits to multi-band data during continuous outbursts, with standstills and gaps as intervals, yield shortest and longest outburst periods of 7.93(1) days and 14.38(3) days, respectively (Fig. \ref{fig:outburst-FT}).

\subsection{Positive Superhumps}\label{sec:Positive Superhumps}

To present the generation and evolution of PSHs in detail, we adopted the analysis method of of recent studies \citep{2023arXiv230911033S,2024arXiv240903011S}, and followed the steps outlined below:
\begin{enumerate}
	\item[(i)] The locally-weighted polynomial regression (LOWESS) \citep{cleveland1979robust} fit with a 2-day span was applied to remove the long-term trends, and the results are shown in Figure \ref{fig:psh1}(a) and (b).
	
	\item[(ii)] Frequency analysis of the residuals (residual$ _{1} $) from the LOWESS fit was performed using the Period04 software \citep{Period04} to obtain the parameters for the orbital signal and its second harmonic. The orbital signal was then removed to eliminate the interference in the PSH calculation using the following formula (Tab. \ref{tab:tab_FT}):
	
	\begin{equation}
		\text{LC$ _\text{orb} $}(t) = Z + \sum \text{Amplitude}_i \cdot \sin(2\pi \cdot (\text{frequency}_i \cdot t + \text{phase}_i))
	\end{equation}
	
	where \(Z\), \(\text{Amplitude}_i\), \(\text{frequency}_i\), and \(\text{phase}_i\) are the fitted intercept, amplitude, frequency, and phase, respectively.
	The residual$ _{2} $ after the orbital signal removal is given by:
	\text{residual$ _{2} $} = \text{residual$ _{1} $} - \text{LC$ _\text{orb} $}.
	The frequency spectra of residual$ _{1} $ and residual$ _{2} $ are shown as black and red lines in the right panel Figure \ref{fig:lc_FT}(e).
	
	\item[(iii)] A segmented frequency analysis was conducted on residual$ _{2} $ with segment intervals of 2 days and 3 days. The results are shown in Figure \ref{fig:2-3_FT} and Table \ref{tab:tab_FT}.
	
	\item[(iv)] The Continuous Wavelet Transform (CWT)\citep{foster1996wavelets} was applied to residual$ _{2} $ as a cross-validation of the segmented frequency analysis. The results are presented in Figure \ref{fig:psh1}(d).
	
	\item[(v)] The significant PSHs in S72 provided an opportunity to determine the parameters of each PSH. We used a linear superposition of Gaussian fits to calculate the peak information of the PSHs (Fig. \ref{fig:psh1}(c)). Based on the fit, the first maxima and the average PSH period from the segmented (3d) analysis of S72, we derived the following epoch:
	
	\begin{equation}
		\text{Maxima} = \text{TJD}3261.4371(23) + E \times 0.22551(33) \, \text{d}
	\end{equation}
	
	Using this epoch formula, an O-C (observed minus calculated) analysis of the PSHs was performed, and the results are shown in Figure \ref{fig:psh1}(g).
	
	\item[(vi)] A parabolic fit was applied to the O-C curve obtained from step (v).
	The results are shown in Figure \ref{fig:S71_O-C}. The O-C curve of AT Cnc suggests a possible Stage B similar to other SU UMa-type DNe. A separate parabolic fit for the Stage B is shown in the black curve in Figure \ref{fig:S71_O-C}.
	
	\item[(vii)] Based on the maxima obtained in step (v) and the two minima before and after the PSH peak, we calculated the half-amplitude of each PSH using the following formula, which serves as a complementary result to the segmented frequency analysis and CWT:
	
	\begin{equation}
		\Delta \text{Amplitude}_{\text{psh}} = \left[  \text{minima}_{\text{before, mag}} + \text{minima}_{\text{after, mag}}  / 2 - \text{maxima}_{\text{mag}}\right] /2
	\end{equation}
	
	Here, \(\text{minima}_{\text{before, mag}}\) and \(\text{minima}_{\text{after, mag}}\) refer to the magnitudes of the minima before and after the PSH maxima, respectively. 
\end{enumerate}

\renewcommand\thefigure{A1}
\begin{figure*}
	\centering
	\includegraphics[width=0.8\columnwidth]{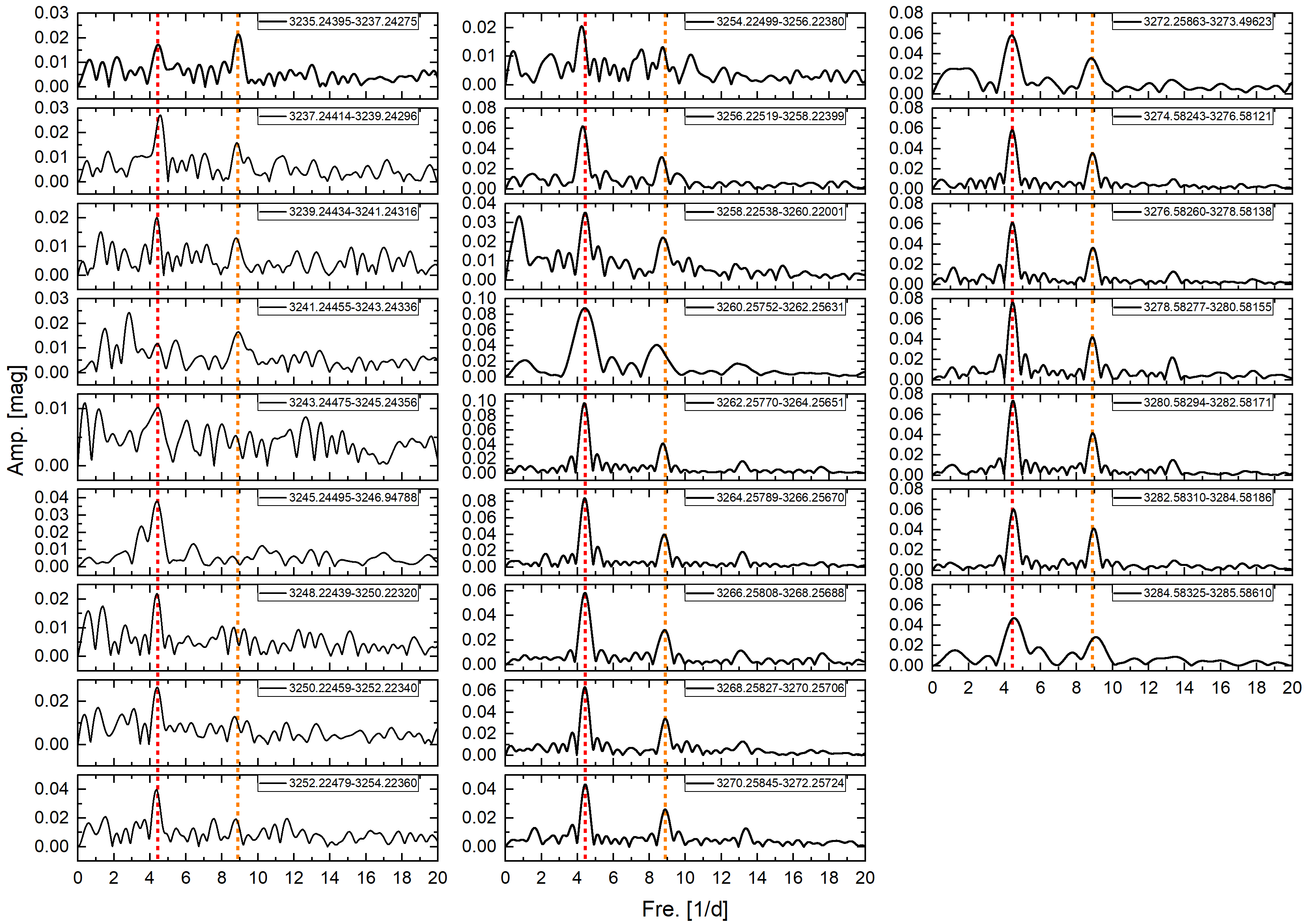}{(a)}\\
	\includegraphics[width=0.8\columnwidth]{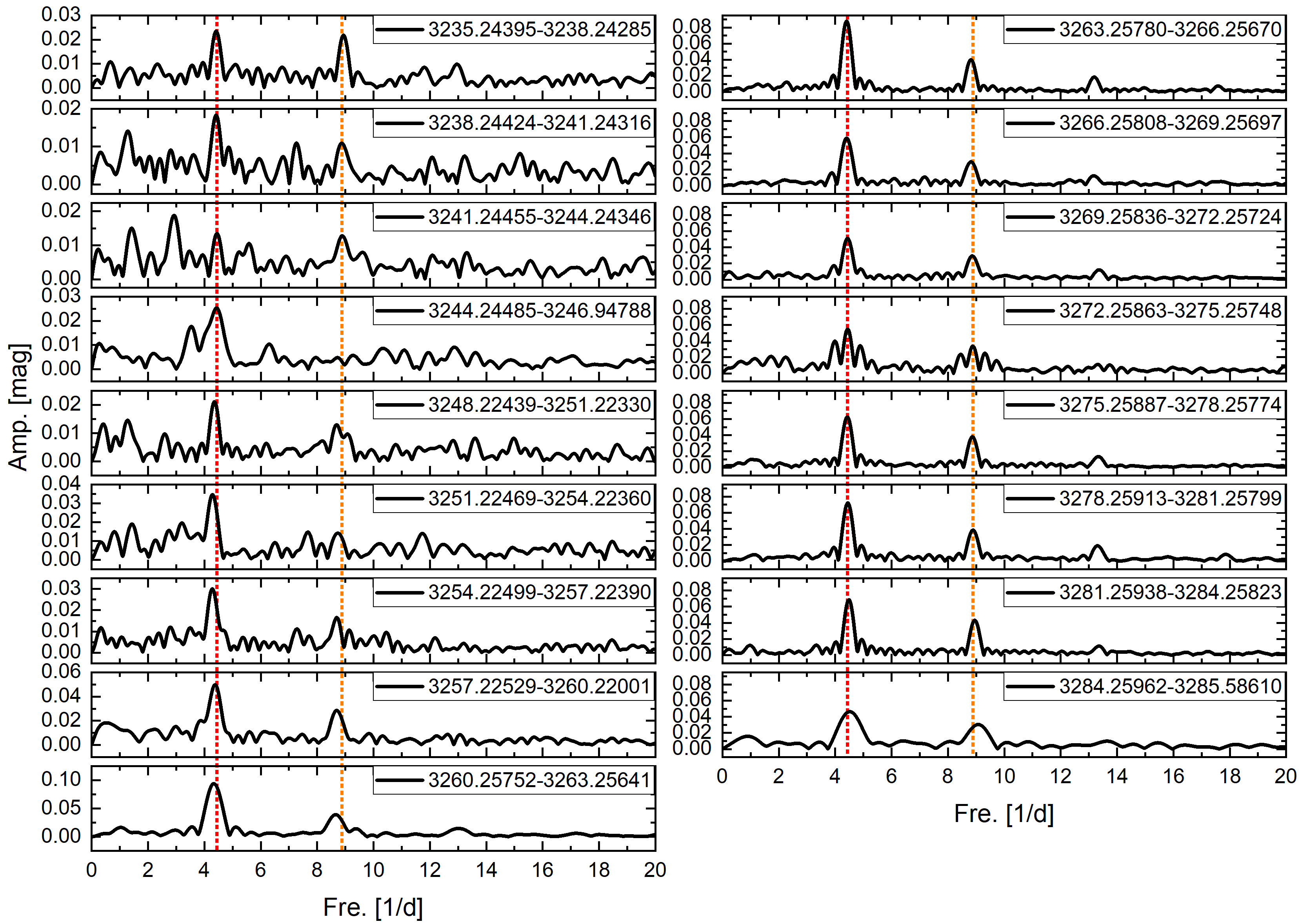}{(b)}
	\caption{Frequency spectra of segmented frequency analysis. (a) and (b) are the results of the segmented for 2 and 3 days analysis. The vertical dashed line in the figure corresponds to the PSH frequency and its second harmonic. The numbers in the legend correspond to the time span of the light curve corresponding to each periodogram, and the unit is TJD.\label{fig:2-3_FT}}
\end{figure*}

\clearpage
\newpage

\renewcommand\thefigure{A2}
\begin{figure*}
	\centering
	\includegraphics[width=0.45\columnwidth]{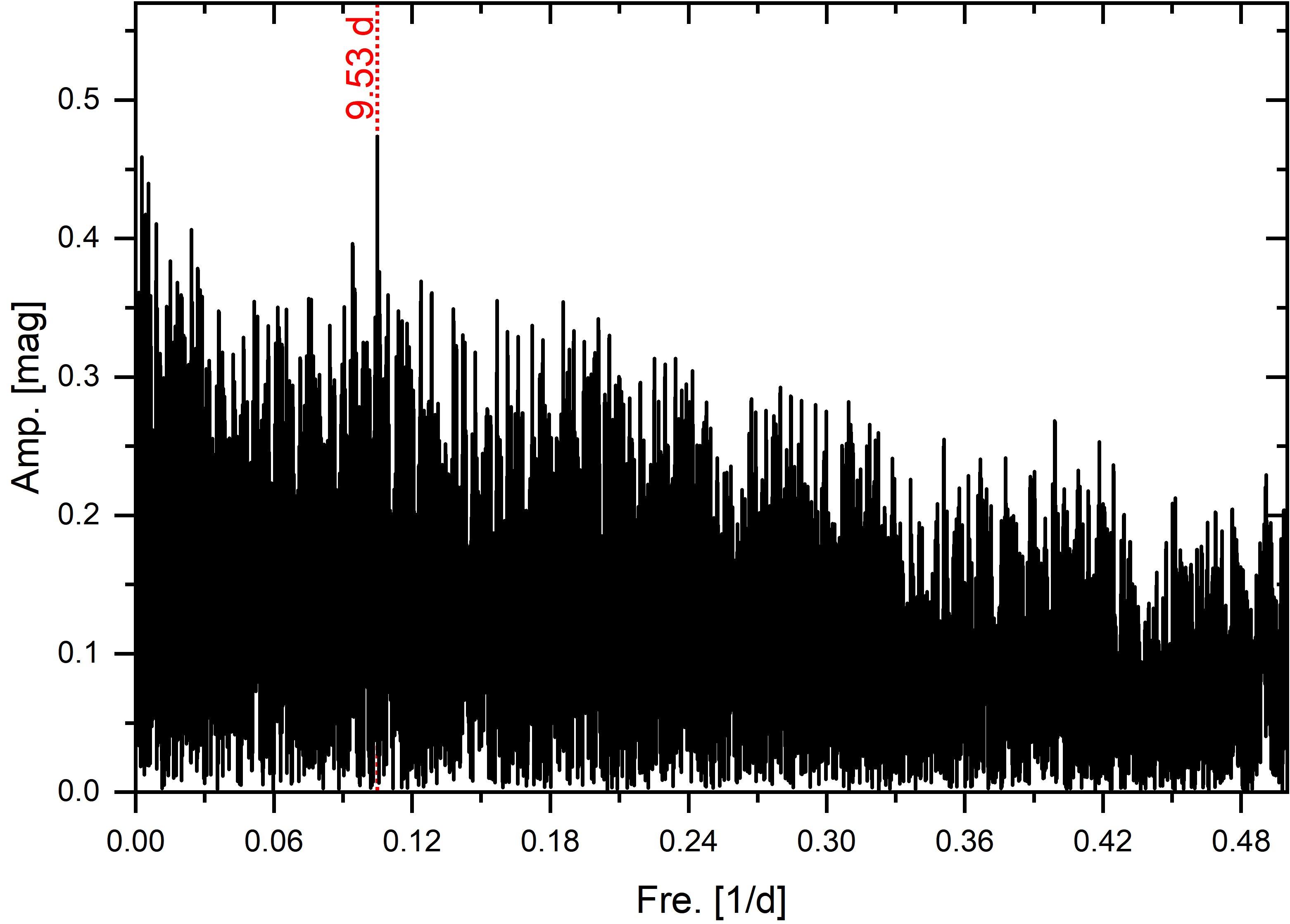}{}
	\includegraphics[width=0.45\columnwidth]{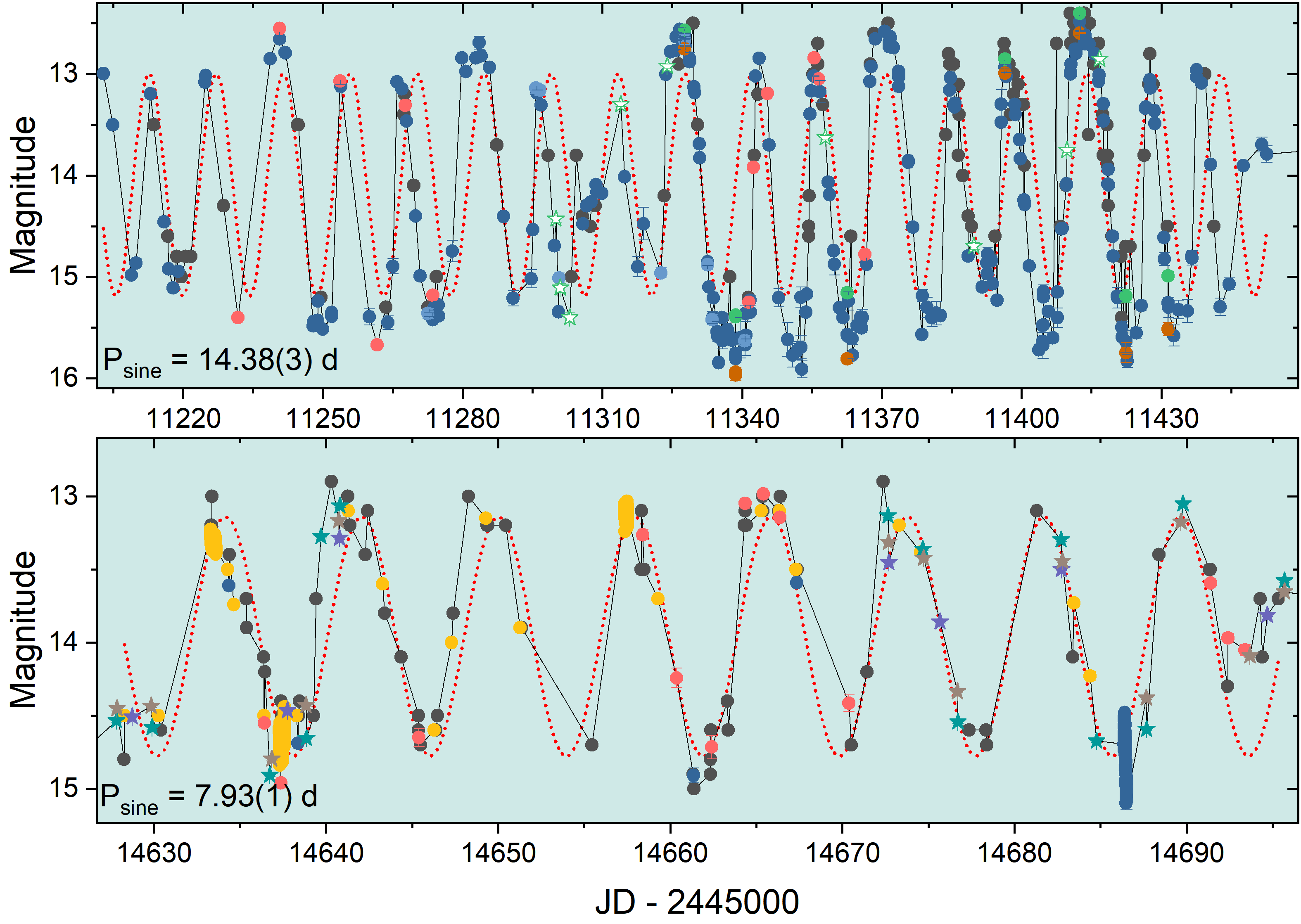}{}
	\caption{Analysis of the AT Cnc outburst periods. Left panel: Frequency spectrum of the V-band light curve of the AT Cnc from AAVSO and ASAS-SN, with the red dashed line corresponding to the peak frequency ; Right panel: light curves corresponding to the longest and shortest outbursts and sine fits. \label{fig:outburst-FT}}
\end{figure*}

\renewcommand\thetable{A3}
\begin{deluxetable*}{lllll}
	\tablecaption{The TESS photometric log for AT Cnc.  \label{tessLOG}}
	\tablewidth{1pt}
	\tabletypesize{\scriptsize}
	\tablehead{	
		\colhead{Sectors}&\colhead{Start time}&\colhead{End time}&\colhead{Start time}&\colhead{End time}\\
	\colhead{}&\colhead{[UT]}&\colhead{[UT]}&\colhead{[TJD]$\rm ^{a} $}&\colhead{[TJD]$\rm ^{a} $}}
	\startdata
	21&2020-01-21T22:33:32.974&2020-02-18T06:50:46.806&1870.44077 &1897.78606\\ 
44&2021-10-12T16:51:01.448&2021-11-05T22:34:27.544&2500.20290 &2524.44140\\ 
45&2021-11-07T05:26:38.228&2021-12-02T03:03:51.738&2525.72763 &2550.62848\\ 
46&2021-12-03T12:05:58.954&2021-12-30T04:56:19.929&2552.00496 &2578.70659\\ 
47&2021-12-31T07:40:23.644&2022-01-27T10:41:03.972&2579.82052 &2606.94599\\ 
71&2023-10-17T17:49:07.811&2023-11-11T09:20:36.938&3235.24325 &3259.89012\\ 
72&2023-11-11T16:24:39.674&2023-12-07T02:03:49.742&3260.18459 &3285.58679\\ 
	\enddata
	\begin{threeparttable}
		\begin{tablenotes}
			\item [a] The units are BJD - 2457000. 
		\end{tablenotes}
	\end{threeparttable}
\end{deluxetable*}

\newpage

\renewcommand\thetable{A4}
\setlength{\tabcolsep}{10pt}
\begin{longtable}{cccccccccc}

	\caption{Frequency analysis results for different TESS sectors and segments.}\label{tab:tab_FT}\\

	
	\toprule
	Sectors/Spans&Name &Fre.&err&Amp.&err&Phase&err\\
	
	-/TJD&&[1/d] &[1/d]&[mag]&[mag]&[rad]&[rad]\\
	\midrule
	\endfirsthead
	\multicolumn{8}{c}
	{{\bfseries{Table \thetable\ }continued from previous page}} \\
	\toprule
	Sectors/Spans&Name &Fre.&err&Amp.&err&Phase&err\\
	
	-/TJD&&[1/d] &[1/d]&[mag]&[mag]&[rad]&[rad]\\
	\midrule
	\endhead
	\midrule
	\multicolumn{8}{c}
	{{ }} \\
	\endfoot
	\endlastfoot
	S21&2orb&9.9154 &0.0014 &0.0063 &0.0004 &0.41 &0.07 \\
	S71&-&0.1978 &0.0001 &0.1119 &0.0007 &0.92 &0.01 \\
	S71&psh&4.3872 &0.0006 &0.0177 &0.0005 &0.47 &0.03 \\
	S71&orb&4.9589 &0.0012 &0.0093 &0.0005 &0.43 &0.06 \\
	S71&2psh&8.6975 &0.0011 &0.0094 &0.0005 &0.48 &0.05 \\
	S72&psh&4.4323 &0.0002 &0.0599 &0.0005 &0.79 &0.01 \\
	S72&2psh&8.8673 &0.0004 &0.0239 &0.0004 &0.07 &0.02 \\
	S72&orb&4.9545 &0.0007 &0.0111 &0.0004 &0.77 &0.03 \\
	S72&2orb&9.9141 &0.0017 &0.0039 &0.0003 &0.38 &0.08 \\
	S44-S47&-&0.1793 &0.0001 &0.0311 &0.0002 &0.77 &0.01 \\
	S44-S47&orb&4.9562 &0.0007 &0.0046 &0.0002 &0.53 &0.03 \\
	S44-S47&2orb&9.9133 &0.0013 &0.0025 &0.0002 &0.65 &0.06 \\
	S71-S72&psh&4.4450 &0.0003 &0.0292 &0.0004 &0.19 &0.01 \\
	S71-S72&2psh&8.8691 &0.0005 &0.0151 &0.0004 &0.18 &0.02 \\
	S71-S72&orb&4.9566 &0.0007 &0.0102 &0.0003 &0.88 &0.03 \\
	S71-S72&2orb&9.9136 &0.0016 &0.0039 &0.0003 &0.02 &0.08 \\
	\midrule
	3235.24395-3237.24275&psh&4.464 &0.027 &0.0176 &0.0017 &0.70 &0.10 \\
	3237.24414-3239.24296&psh&4.567 &0.018 &0.0271 &0.0018 &0.12 &0.07 \\
	3239.24434-3241.24316&psh&4.382 &0.022 &0.0200 &0.0016 &0.33 &0.08 \\
	3241.24455-3243.24336&psh&4.305 &0.038 &0.0115 &0.0016 &0.98 &0.14 \\
	3243.24475-3245.24356&psh&4.428 &0.040 &0.0100 &0.0015 &0.17 &0.15 \\
	3245.24495-3246.94788&psh&4.387 &0.015 &0.0380 &0.0021 &0.26 &0.05 \\
	3248.22439-3250.22320&psh&4.388 &0.021 &0.0218 &0.0016 &0.14 &0.08 \\
	3250.22459-3252.22340&psh&4.399 &0.019 &0.0265 &0.0018 &0.32 &0.07 \\
	3252.22479-3254.22360&psh&4.357 &0.017 &0.0397 &0.0024 &0.72 &0.06 \\
	3254.22499-3256.22380&psh&4.234 &0.019 &0.0205 &0.0014 &0.27 &0.07 \\
	3256.22519-3258.22399&psh&4.295 &0.009 &0.0620 &0.0021 &0.71 &0.03 \\
	3258.22538-3260.22001&psh&4.428 &0.016 &0.0353 &0.0020 &0.37 &0.06 \\
	3260.25752-3262.25631&psh&4.408 &0.008 &0.0882 &0.0027 &0.21 &0.03 \\
	3262.25770-3264.25651&psh&4.383 &0.004 &0.0976 &0.0016 &0.71 &0.02 \\
	3264.25789-3266.25670&psh&4.403 &0.005 &0.0841 &0.0015 &0.45 &0.02 \\
	3266.25808-3268.25688&psh&4.416 &0.007 &0.0584 &0.0015 &0.01 &0.03 \\
	3268.25827-3270.25706&psh&4.409 &0.007 &0.0632 &0.0017 &0.88 &0.03 \\
	3270.25845-3272.25724&psh&4.431 &0.010 &0.0435 &0.0015 &0.97 &0.03 \\
	3272.25863-3273.49623&psh&4.394 &0.014 &0.0582 &0.0029 &0.06 &0.05 \\
	3274.58243-3276.58121&psh&4.423 &0.008 &0.0582 &0.0016 &0.20 &0.03 \\
	3276.58260-3278.58138&psh&4.434 &0.007 &0.0610 &0.0015 &0.17 &0.03 \\
	3278.58277-3280.58155&psh&4.449 &0.007 &0.0763 &0.0019 &0.99 &0.02 \\
	3280.58294-3282.58171&psh&4.464 &0.006 &0.0735 &0.0015 &0.76 &0.02 \\
	3282.58310-3284.58186&psh&4.493 &0.008 &0.0601 &0.0017 &0.57 &0.03 \\
	3284.58325-3285.58610&psh&4.532 &0.013 &0.0470 &0.0023 &0.48 &0.05 \\
	\midrule
	3235.24395-3238.24285&psh&4.417 &0.017 &0.0235 &0.0015 &0.79 &0.06 \\
	3238.24424-3241.24316&psh&4.417 &0.020 &0.0183 &0.0014 &0.92 &0.07 \\
	3241.24455-3244.24346&psh&4.452 &0.027 &0.0138 &0.0014 &0.37 &0.10 \\
	3244.24485-3246.94788&psh&4.434 &0.017 &0.0253 &0.0015 &0.68 &0.06 \\
	3248.22439-3251.22330&psh&4.357 &0.017 &0.0211 &0.0013 &0.86 &0.06 \\
	3251.22469-3254.22360&psh&4.288 &0.014 &0.0346 &0.0018 &0.21 &0.05 \\
	3254.22499-3257.22390&psh&4.282 &0.012 &0.0300 &0.0013 &0.02 &0.04 \\
	3257.22529-3260.22001&psh&4.380 &0.010 &0.0497 &0.0018 &0.80 &0.04 \\
	3260.25752-3263.25641&psh&4.336 &0.005 &0.0940 &0.0017 &0.06 &0.02 \\
	3263.25780-3266.25670&psh&4.406 &0.004 &0.0877 &0.0011 &0.65 &0.01 \\
	3266.25808-3269.25697&psh&4.408 &0.006 &0.0590 &0.0012 &0.15 &0.02 \\
	3269.25836-3272.25724&psh&4.439 &0.007 &0.0510 &0.0013 &0.80 &0.02 \\
	3272.25863-3275.25748&psh&4.443 &0.011 &0.0548 &0.0021 &0.69 &0.04 \\
	3275.25887-3278.25774&psh&4.430 &0.005 &0.0626 &0.0013 &0.27 &0.02 \\
	3278.25913-3281.25799&psh&4.448 &0.005 &0.0721 &0.0013 &0.27 &0.02 \\
	3281.25938-3284.25823&psh&4.486 &0.005 &0.0686 &0.0014 &0.55 &0.02 \\
	3284.25962-3285.58610&psh&4.519 &0.012 &0.0463 &0.0021 &0.19 &0.04 \\
	\bottomrule
\end{longtable}

\newpage
\bibliography{sample631}{}

\begin{thebibliography}{}
\expandafter\ifx\csname natexlab\endcsname\relax\def\natexlab#1{#1}\fi
\providecommand{\url}[1]{\href{#1}{#1}}
\providecommand{\dodoi}[1]{doi:~\href{http://doi.org/#1}{\nolinkurl{#1}}}
\providecommand{\doeprint}[1]{\href{http://ascl.net/#1}{\nolinkurl{http://ascl.net/#1}}}
\providecommand{\doarXiv}[1]{\href{https://arxiv.org/abs/#1}{\nolinkurl{https://arxiv.org/abs/#1}}}

\bibitem[{{Barrett} {et~al.}(1988){Barrett}, {O'Donoghue}, \&
  {Warner}}]{Barrett1988MNRAS}
{Barrett}, P., {O'Donoghue}, D., \& {Warner}, B. 1988, \mnras, 233, 759,
  \dodoi{10.1093/mnras/233.4.759}

\bibitem[{{Bellm} {et~al.}(2019){Bellm}, {Kulkarni}, {Graham}, {Dekany},
  {Smith}, {Riddle}, {Masci}, {Helou}, {Prince}, {Adams}, {Barbarino},
  {Barlow}, {Bauer}, {Beck}, {Belicki}, {Biswas}, {Blagorodnova}, {Bodewits},
  {Bolin}, {Brinnel}, {Brooke}, {Bue}, {Bulla}, {Burruss}, {Cenko}, {Chang},
  {Connolly}, {Coughlin}, {Cromer}, {Cunningham}, {De}, {Delacroix}, {Desai},
  {Duev}, {Eadie}, {Farnham}, {Feeney}, {Feindt}, {Flynn}, {Franckowiak},
  {Frederick}, {Fremling}, {Gal-Yam}, {Gezari}, {Giomi}, {Goldstein},
  {Golkhou}, {Goobar}, {Groom}, {Hacopians}, {Hale}, {Henning}, {Ho}, {Hover},
  {Howell}, {Hung}, {Huppenkothen}, {Imel}, {Ip}, {Ivezi{\'c}}, {Jackson},
  {Jones}, {Juric}, {Kasliwal}, {Kaspi}, {Kaye}, {Kelley}, {Kowalski},
  {Kramer}, {Kupfer}, {Landry}, {Laher}, {Lee}, {Lin}, {Lin}, {Lunnan},
  {Giomi}, {Mahabal}, {Mao}, {Miller}, {Monkewitz}, {Murphy}, {Ngeow},
  {Nordin}, {Nugent}, {Ofek}, {Patterson}, {Penprase}, {Porter}, {Rauch},
  {Rebbapragada}, {Reiley}, {Rigault}, {Rodriguez}, {van Roestel}, {Rusholme},
  {van Santen}, {Schulze}, {Shupe}, {Singer}, {Soumagnac}, {Stein}, {Surace},
  {Sollerman}, {Szkody}, {Taddia}, {Terek}, {Van Sistine}, {van Velzen},
  {Vestrand}, {Walters}, {Ward}, {Ye}, {Yu}, {Yan}, \& {Zolkower}}]{2019ZTF}
{Bellm}, E.~C., {Kulkarni}, S.~R., {Graham}, M.~J., {et~al.} 2019, \pasp, 131,
  018002, \dodoi{10.1088/1538-3873/aaecbe}

\bibitem[{{Bruch}(2023)}]{BruchIII}
{Bruch}, A. 2023, \mnras, 525, 1953, \dodoi{10.1093/mnras/stad2089}

\bibitem[{{Bruch} {et~al.}(2019){Bruch}, {Boardman}, {Cook}, {Cook}, {Dvorak},
  {Jones}, {Rock}, {Stone}, \& {Ulowetz}}]{2019NewA...67...22B}
{Bruch}, A., {Boardman}, J., {Cook}, L.~M., {et~al.} 2019, \na, 67, 22,
  \dodoi{10.1016/j.newast.2018.09.001}

\bibitem[{{Butters} {et~al.}(2010){Butters}, {West}, {Anderson}, {Collier
  Cameron}, {Clarkson}, {Enoch}, {Haswell}, {Hellier}, {Horne}, {Joshi},
  {Kane}, {Lister}, {Maxted}, {Parley}, {Pollacco}, {Smalley}, {Street},
  {Todd}, {Wheatley}, \& {Wilson}}]{2010A&A...520L..10B}
{Butters}, O.~W., {West}, R.~G., {Anderson}, D.~R., {et~al.} 2010, \aap, 520,
  L10, \dodoi{10.1051/0004-6361/201015655}

\bibitem[{Cleveland(1979)}]{cleveland1979robust}
Cleveland, W.~S. 1979, Journal of the American statistical association, 74, 829

\bibitem[{{Court} {et~al.}(2019){Court}, {Scaringi}, {Rappaport}, {Zhan},
  {Littlefield}, {Castro Segura}, {Knigge}, {Maccarone}, {Kennedy}, {Szkody},
  \& {Garnavich}}]{2019MNRAS.488.4149C}
{Court}, J.~M.~C., {Scaringi}, S., {Rappaport}, S., {et~al.} 2019, \mnras, 488,
  4149, \dodoi{10.1093/mnras/stz2015}

\bibitem[{Dubus {et~al.}(2018)Dubus, Otulakowska-Hypka, \&
  Lasota}]{dubus2018testing}
Dubus, G., Otulakowska-Hypka, M., \& Lasota, J.-P. 2018, \aap, 617, A26,
  \dodoi{10.1051/0004-6361/201833372}

\bibitem[{Foster(1996)}]{foster1996wavelets}
Foster, G. 1996, \aj, 112, 1709, \dodoi{10.1086/118137}

\bibitem[{{G{\"o}tz}(1991)}]{Gotz1991}
{G{\"o}tz}, W. 1991, Zentralinstitut fuer Astrophysik Sternwarte Sonneberg
  Mitteilungen ueber Veraenderliche Sterne, 12, 111

\bibitem[{{Hameury}(2020)}]{2020AdSpR..66.1004H}
{Hameury}, J.~M. 2020, Advances in Space Research, 66, 1004,
  \dodoi{10.1016/j.asr.2019.10.022}

\bibitem[{{Hameury} \& {Lasota}(2014)}]{2014mass-transferoutbursts}
{Hameury}, J.~M., \& {Lasota}, J.~P. 2014, \aap, 569, A48,
  \dodoi{10.1051/0004-6361/201424535}

\bibitem[{{Jayasinghe} {et~al.}(2019){Jayasinghe}, {Stanek}, {Kochanek},
  {Shappee}, {Holoien}, {Thompson}, {Prieto}, {Dong}, {Pawlak}, {Pejcha},
  {Shields}, {Pojmanski}, {Otero}, {Britt}, \& {Will}}]{2019MNRAS.486.1907J}
{Jayasinghe}, T., {Stanek}, K.~Z., {Kochanek}, C.~S., {et~al.} 2019, \mnras,
  486, 1907, \dodoi{10.1093/mnras/stz844}

\bibitem[{{Jenkins} {et~al.}(2016){Jenkins}, {Twicken}, {McCauliff},
  {Campbell}, {Sanderfer}, {Lung}, {Mansouri-Samani}, {Girouard}, {Tenenbaum},
  {Klaus}, {Smith}, {Caldwell}, {Chacon}, {Henze}, {Heiges}, {Latham},
  {Morgan}, {Swade}, {Rinehart}, \& {Vanderspek}}]{2016SPIE.9913E..3EJ}
{Jenkins}, J.~M., {Twicken}, J.~D., {McCauliff}, S., {et~al.} 2016, in Society
  of Photo-Optical Instrumentation Engineers (SPIE) Conference Series, Vol.
  9913, Software and Cyberinfrastructure for Astronomy IV, ed. G.~{Chiozzi} \&
  J.~C. {Guzman}, 99133E, \dodoi{10.1117/12.2233418}

\bibitem[{{Kato}(2019)}]{Kato2019PASJ}
{Kato}, T. 2019, \pasj, 71, 20, \dodoi{10.1093/pasj/psy138}

\bibitem[{{Kato} {et~al.}(2008){Kato}, {Maehara}, \&
  {Monard}}]{2008PASJ...60L..23K}
{Kato}, T., {Maehara}, H., \& {Monard}, B. 2008, \pasj, 60, L23,
  \dodoi{10.1093/pasj/60.4.L23}

\bibitem[{{Kato} {et~al.}(2001){Kato}, {Stubbings}, {Reszelski}, {Muyllaert},
  {Simonsen}, {Poyner}, {Dubovsky}, {Pearce}, {Kinnunen}, \&
  {Maehara}}]{2001IBVS.5099....1K}
{Kato}, T., {Stubbings}, R., {Reszelski}, M., {et~al.} 2001, Information
  Bulletin on Variable Stars, 5099, 1

\bibitem[{{Kato} {et~al.}(2009){Kato}, {Imada}, {Uemura}, {Nogami}, {Maehara},
  {Ishioka}, {Baba}, {Matsumoto}, {Iwamatsu}, {Kubota}, {Sugiyasu}, {Soejima},
  {Moritani}, {Ohshima}, {Ohashi}, {Tanaka}, {Sasada}, {Arai}, {Nakajima},
  {Kiyota}, {Tanabe}, {Imamura}, {Kunitomi}, {Kunihiro}, {Taguchi}, {Koizumi},
  {Yamada}, {Nishi}, {Kida}, {Tanaka}, {Ueoka}, {Yasui}, {Maruoka}, {Henden},
  {Oksanen}, {Moilanen}, {Tikkanen}, {Aho}, {Monard}, {Itoh}, {Dubovsky},
  {Kudzej}, {Dancikova}, {Vanmunster}, {Pietz}, {Bolt}, {Boyd}, {Nelson},
  {Krajci}, {Cook}, {Torii}, {Starkey}, {Shears}, {Jensen}, {Masi}, {Hynek},
  {Nov{\'a}k}, {Koci{\'a}n}, {Kr{\'a}l}, {Ku{\v{c}}{\'a}kov{\'a}}, {Kolasa},
  {{\v{S}}tastn{\'y}}, {Staels}, {Miller}, {Sano}, {Ponthi{\`e}re},
  {Miyashita}, {Crawford}, {Brady}, {Santallo}, {Richards}, {Martin},
  {Buczynski}, {Richmond}, {Kern}, {Davis}, {Crabtree}, {Beaulieu}, {Davis},
  {Aggleton}, {Morelle}, {Pavlenko}, {Andreev}, {Baklanov}, {Koppelman},
  {Billings}, {Urban{\v{c}}ok}, {{\"O}gmen}, {Heathcote}, {Gomez}, {Voloshina},
  {Retter}, {Mularczyk}, {Z{\l}oczewski}, {Olech}, {Kedzierski}, {Pickard},
  {Stockdale}, {Virtanen}, {Morikawa}, {Hambsch}, {Garradd}, {Gualdoni},
  {Geary}, {Omodaka}, {Sakai}, {Michel}, {C{\'a}rdenas}, {Gazeas}, {Niarchos},
  {Yushchenko}, {Mallia}, {Fiaschi}, {Good}, {Walker}, {James}, {Douzu},
  {Julian}, {Butterworth}, {Shugarov}, {Volkov}, {Chochol}, {Katysheva},
  {Rosenbush}, {Khramtsova}, {Kehusmaa}, {Reszelski}, {Bedient}, {Liller},
  {Pojma{\'n}ski}, {Simonsen}, {Stubbings}, {Schmeer}, {Muyllaert}, {Kinnunen},
  {Poyner}, {Ripero}, \& {Kriebel}}]{2009PASJ...61S.395K}
{Kato}, T., {Imada}, A., {Uemura}, M., {et~al.} 2009, \pasj, 61, S395,
  \dodoi{10.1093/pasj/61.sp2.S395}

\bibitem[{Kato {et~al.}(2015)Kato, Hambsch, Dubovsky, Kudzej, Monard, Miller,
  Itoh, Kiyota, Masumoto, Fukushima, {et~al.}}]{Kato2015PASJ}
Kato, T., Hambsch, F.-J., Dubovsky, P.~A., {et~al.} 2015, \pasj, 67, 105

\bibitem[{{Kato} {et~al.}(2019){Kato}, {Pavlenko}, {Pit}, {Antonyuk},
  {Antonyuk}, {Babina}, {Baklanov}, {Sosnovskij}, {Belan}, {Maeda}, {Sugiura},
  {Sumiya}, {Matsumoto}, {Ito}, {Nikai}, {Kojiguchi}, {Matsumoto}, {Dubovsky},
  {Kudzej}, {Medulka}, {Wakamatsu}, {Ohnishi}, {Seki}, {Isogai}, {Simon},
  {Romanjuk}, {Baransky}, {Sergeev}, {Godunova}, {Izviekova}, {Kozlov},
  {Sklyanov}, {Zhuchkov}, {Gutaev}, {Ponomarenko}, {Vasylenko}, {Miller},
  {Kasai}, {Dvorak}, {Menzies}, {de Miguel}, {Brincat}, \&
  {Pickard}}]{2019PASJ...71L...1K}
{Kato}, T., {Pavlenko}, E.~P., {Pit}, N.~V., {et~al.} 2019, \pasj, 71, L1,
  \dodoi{10.1093/pasj/psz007}

\bibitem[{{Kato} {et~al.}(2021){Kato}, {Tampo}, {Kojiguchi}, {Shibata}, {Ito},
  {Isogai}, {Itoh}, {Hambsch}, {Monard}, {Kiyota}, {Vanmunster}, {Sosnovskij},
  {Pavlenko}, {Dubovsky}, {Kudzej}, \& {Medulka}}]{2021PASJ...73.1280K}
{Kato}, T., {Tampo}, Y., {Kojiguchi}, N., {et~al.} 2021, \pasj, 73, 1280,
  \dodoi{10.1093/pasj/psab074}

\bibitem[{Kimura {et~al.}(2020b)Kimura, Osaki, Kato, \&
  Mineshige}]{kimura2020thermal}
Kimura, M., Osaki, Y., Kato, T., \& Mineshige, S. 2020b, \pasj, 72, 22,
  \dodoi{10.1093/pasj/psz144}

\bibitem[{{Knigge} {et~al.}(2000){Knigge}, {King}, \&
  {Patterson}}]{2000A&A...364L..75K}
{Knigge}, C., {King}, A.~R., \& {Patterson}, J. 2000, \aap, 364, L75,
  \dodoi{10.48550/arXiv.astro-ph/0011304}

\bibitem[{{Kozhevnikov}(2004)}]{2004A&A...419.1035K}
{Kozhevnikov}, V.~P. 2004, \aap, 419, 1035, \dodoi{10.1051/0004-6361:20035600}

\bibitem[{Lasota(2001)}]{lasota2001disc}
Lasota, J.-P. 2001, New Astronomy Reviews, 45, 449,
  \dodoi{10.1016/S1387-6473(01)00112-9}

\bibitem[{{Lenz} \& {Breger}(2005)}]{Period04}
{Lenz}, P., \& {Breger}, M. 2005, Communications in Asteroseismology, 146, 53,
  \dodoi{10.1553/cia146s53}

\bibitem[{{Lubow}(1991)}]{1991ApJ...381..259L}
{Lubow}, S.~H. 1991, \apj, 381, 259, \dodoi{10.1086/170647}

\bibitem[{{Meyer} \& {Meyer-Hofmeister}(1983)}]{Meyer1983A&A}
{Meyer}, F., \& {Meyer-Hofmeister}, E. 1983, \aap, 121, 29

\bibitem[{{Murray} {et~al.}(2000){Murray}, {Warner}, \&
  {Wickramasinghe}}]{2000NewAR..44...51M}
{Murray}, J., {Warner}, B., \& {Wickramasinghe}, D. 2000, \nar, 44, 51,
  \dodoi{10.1016/S1387-6473(00)00013-0}

\bibitem[{{Murray}(2000)}]{2000MNRAS.314L...1M}
{Murray}, J.~R. 2000, \mnras, 314, L1, \dodoi{10.1046/j.1365-8711.2000.03424.x}

\bibitem[{{Nogami} {et~al.}(1999){Nogami}, {Masuda}, {Kato}, \&
  {Hirata}}]{1999PASJ...51..115N}
{Nogami}, D., {Masuda}, S., {Kato}, T., \& {Hirata}, R. 1999, \pasj, 51, 115,
  \dodoi{10.1093/pasj/51.1.115}

\bibitem[{{Osaki}(1974)}]{1974PASJOsaki}
{Osaki}, Y. 1974, \pasj, 26, 429

\bibitem[{Osaki(1985)}]{osaki1985irradiation}
Osaki, Y. 1985, \aap, 144, 369

\bibitem[{{Osaki}(1989)}]{1989PASJ...41.1005O}
{Osaki}, Y. 1989, Publications of the Astronomical Society of Japan, 41, 1005

\bibitem[{{Osaki}(2005)}]{2005PJABOsaki}
---. 2005, Proceedings of the Japan Academy, Series B, 81, 291,
  \dodoi{10.2183/pjab.81.291}

\bibitem[{{Osaki} \& {Kato}(2013)}]{Osaki2013b}
{Osaki}, Y., \& {Kato}, T. 2013, \pasj, 65, 95, \dodoi{10.1093/pasj/65.5.95}

\bibitem[{{Osaki} \& {Meyer}(2003)}]{2003A&A...401..325O}
{Osaki}, Y., \& {Meyer}, F. 2003, \aap, 401, 325,
  \dodoi{10.1051/0004-6361:20030115}

\bibitem[{{Paczynski}(1977)}]{Paczynski1977}
{Paczynski}, B. 1977, Astrophysical Journal, 216, 822, \dodoi{10.1086/155526}

\bibitem[{{Patterson} {et~al.}(2005){Patterson}, {Kemp}, {Harvey}, {Fried},
  {Rea}, {Monard}, {Cook}, {Skillman}, {Vanmunster}, {Bolt}, {Armstrong},
  {McCormick}, {Krajci}, {Jensen}, {Gunn}, {Butterworth}, {Foote}, {Bos},
  {Masi}, \& {Warhurst}}]{2005PASP..117.1204P}
{Patterson}, J., {Kemp}, J., {Harvey}, D.~A., {et~al.} 2005, \pasp, 117, 1204,
  \dodoi{10.1086/447771}

\bibitem[{Ricker {et~al.}(2015)Ricker, Winn, \& Vanderspek}]{Ricker2015journal}
Ricker, G., Winn, J., \& Vanderspek, R. 2015, JATIS, 1, 014003,
  \dodoi{10.1117/1.JATIS.1.1.014003}

\bibitem[{{Shara} {et~al.}(2017){Shara}, {Drissen}, {Martin}, {Alarie}, \&
  {Stephenson}}]{2017MNRAS.465..739S}
{Shara}, M.~M., {Drissen}, L., {Martin}, T., {Alarie}, A., \& {Stephenson},
  F.~R. 2017, \mnras, 465, 739, \dodoi{10.1093/mnras/stw2753}

\bibitem[{{Shara} {et~al.}(2012){Shara}, {Mizusawa}, {Wehinger}, {Zurek},
  {Martin}, {Neill}, {Forster}, \& {Seibert}}]{2012ApJ...758..121S}
{Shara}, M.~M., {Mizusawa}, T., {Wehinger}, P., {et~al.} 2012, \apj, 758, 121,
  \dodoi{10.1088/0004-637X/758/2/121}

\bibitem[{{Simonsen}(2011)}]{Simonsen2011AAS}
{Simonsen}, M. 2011, in American Astronomical Society Meeting Abstracts, Vol.
  218, American Astronomical Society Meeting Abstracts \#218, 103.02

\bibitem[{{Smak}(1983)}]{Smak1983ApJ}
{Smak}, J. 1983, \apj, 272, 234, \dodoi{10.1086/161284}

\bibitem[{{Smak}(1995)}]{1995AcA....45..355S}
---. 1995, \actaa, 45, 355

\bibitem[{{Sun} {et~al.}(2023c){Sun}, {Qian}, \& {Li}}]{2023arXiv230905891S}
{Sun}, Q.-B., {Qian}, S.-B., \& {Li}, M.-Y. 2023c, \apj, 955, 135,
  \dodoi{10.3847/1538-4357/ace183}

\bibitem[{Sun {et~al.}(2024d)Sun, Qian, Zhu, Li, Li, Li, \&
  Li}]{2024arXiv240903011S}
Sun, Q.-B., Qian, S.-B., Zhu, L.-Y., {et~al.} 2024d, \apj, 976, 107,
  \dodoi{10.3847/1538-4357/ad8446}

\bibitem[{{Sun} {et~al.}(2024c){Sun}, {Qian}, {Zhu}, {Li}, {Li}, \&
  {Li}}]{2024arXiv240704913S}
{Sun}, Q.-B., {Qian}, S.-B., {Zhu}, L.-Y., {et~al.} 2024c, \apj, 974, 132,
  \dodoi{10.3847/1538-4357/ad6f05}

\bibitem[{{Sun} {et~al.}(2023b){Sun}, {Qian}, {Zhu}, {Liao}, {Zhao}, {Li},
  {Shi}, \& {Li}}]{Sun2023arXiv}
---. 2023b, \mnras, 526, 3730, \dodoi{10.1093/mnras/stad1880}

\bibitem[{{Sun} {et~al.}(2024a){Sun}, {Qian}, {Zhu}, {Liao}, {Zhao}, {Li},
  {Shi}, \& {Li}}]{2023arXiv230911033S}
---. 2024a, \apj, 962, 123, \dodoi{10.3847/1538-4357/ad0f1c}

\bibitem[{{Sun} {et~al.}(2024b){Sun}, {Qian}, {Zhu}, {Liao}, {Zhao}, {Li},
  {Shi}, \& {Li}}]{Sun2024ApJ}
---. 2024b, \apj, 966, 83, \dodoi{10.3847/1538-4357/ad2fc2}

\bibitem[{{Sun} {et~al.}(2023a){Sun}, {Qian}, {Zhu}, {Dong}, {Zhi}, {Liao},
  {Zhao}, {Han}, {Liu}, {Zang}, {Li}, \& {Shi}}]{Sun2023MNRAS}
---. 2023a, \mnras, 518, 3901, \dodoi{10.1093/mnras/stac3272}

\bibitem[{{Wakamatsu} {et~al.}(2021){Wakamatsu}, {Thorstensen}, {Kojiguchi},
  {Isogai}, {Kimura}, {Ohnishi}, {Kato}, {Itoh}, {Sugiura}, {Sumiya},
  {Matsumoto}, {Ito}, {Nikai}, {Akitaya}, {Ishioka}, {Oide}, {Kanai}, {Uzawa},
  {Oasa}, {Tordai}, {Vanmunster}, {Shugarov}, {Yamanaka}, {Sasada}, {Takagi},
  {Nishinaka}, {Yamazaki}, {Otsubo}, {Nakaoka}, {Murata}, {Ohsawa}, {Morita},
  {Ichiki}, {Dufoer}, {Mizutani}, {Horiuchi}, {Tozuka}, {Takayama}, {Ohshima},
  {Saito}, {Dubovsky}, {Stone}, {Miller}, \& {Nogami}}]{2021PASJ...73.1209W}
{Wakamatsu}, Y., {Thorstensen}, J.~R., {Kojiguchi}, N., {et~al.} 2021, \pasj,
  73, 1209, \dodoi{10.1093/pasj/psab003}

\bibitem[{Warner(1995)}]{warner1995cat}
Warner, B. 1995, Cambridge University Press, 28

\bibitem[{{Wei} \& {Shengbang}(2023)}]{2023ApJ...954..135W}
{Wei}, L., \& {Shengbang}, Q. 2023, \apj, 954, 135,
  \dodoi{10.3847/1538-4357/acebdf}

\bibitem[{{Whitehurst}(1988)}]{1988MNRAS.232...35W}
{Whitehurst}, R. 1988, \mnras, 232, 35, \dodoi{10.1093/mnras/232.1.35}

\bibitem[{Whitehurst \& King(1991)}]{whitehurst1991superhumps}
Whitehurst, R., \& King, A. 1991, \mnras, 249, 25,
  \dodoi{10.1093/mnras/249.1.25}

\bibitem[{Wood {et~al.}(2009)Wood, Thomas, \& Simpson}]{wood2009sph}
Wood, M.~A., Thomas, D.~M., \& Simpson, J.~C. 2009, \mnras, 398, 2110,
  \dodoi{10.1111/j.1365-2966.2009.15252.x}

\end{thebibliography}
\bibliographystyle{aasjournal}

\end{document}